\documentclass[aps,twocolumn,amsmath,amssymb,superscriptaddress,preprintnumbers]{revtex4-2}
\usepackage[pdftex]{graphicx}
\usepackage[outdir=./]{epstopdf}
\usepackage{color,hyperref}
 
\newcommand{\br}{\boldsymbol{r}}
\newcommand{\bk}{\boldsymbol{k}}

\newcommand{\im}{\mathrm{Im}}

\newcommand{\modify}[1]{{#1}}

\begin{document}

\title{Long-range spin transport on the surface of topological Dirac semimetal}
\author{Yasufumi Araki}
\affiliation{Advanced Science Research Center, Japan Atomic Energy Agency, Tokai 319-1195, Japan}
\author{Takahiro Misawa}
\affiliation{
Beijing Academy of Quantum Information Sciences,Haidian District, Beijing 100193, China}
\author{Kentaro Nomura}
\affiliation{Institute for Materials Research, Tohoku University,
Sendai 980-8577, Japan}
\affiliation{Center for Spintronics Research Network, Tohoku University, Sendai 980-8577, Japan}

\begin{abstract}
    We theoretically propose the long-range spin transport mediated by the gapless surface states of topological Dirac semimetal (TDSM).
    \modify{
        Low-dissipation spin current is a building block of next-generation spintronics devices. 
        While conduction electrons in metals and spin waves in ferromagnetic insulators (FMIs) are the major carriers of spin current,
        their propagation length is inevitably limited due to the Joule heating or the Gilbert damping.
        In order to suppress dissipation and realize long-range spin transport,
        we here make use of the spin-helical surface states of TDSMs, such as $\mathrm{Cd_3 As_2}$ and $\mathrm{Na_3 Bi}$,
        which are robust against disorder.
        Based on a junction of two FMIs connected by a TDSM,
        we demonstrate that the magnetization dynamics in one FMI induces a spin current on the TDSM surface flowing to the other FMI.
        By both the analytical transport theory on the surface and the numerical simulation of real-time evolution in the bulk,
        we find that the induced spin current takes a universal semi-quantized value that is insensitive to the microscopic coupling structure between the FMI and the TDSM.
        We show that this surface spin current is robust against disorder over a long range,
        which indicates that the TDSM surface serves as a promising system for realizing spintronics devices.
        }
\end{abstract}

\preprint{}

\maketitle

\section{Introduction}

Transmission of signals over a long distance is \modify{essential}
in designing integrated information devices.
While charge current in normal metals is inevitably subject to the Joule heating,
spin current is now intensely studied to realize a long-range transmission with less dissipation,
in the context of spintronics \cite{Dyakonov_2008,Takahashi_2008,Maekawa_spin_current}.
Spin current, namely the flow of spin angular momentum,
is carried by various types of quasiparticle excitations in materials.
In metals, spin current is carried by conduction electrons \cite{Johnson_1985}.
Electron spin current can be generated by current injection from magnetic metals \cite{Fert_1969},
spin pumping from magnetic materials with magnetization dynamics \cite{Silsbee_1979,Tserkovnyak_2002,Tserkovnyak_2005},
\modify{
    the spin Hall effect \cite{Dyakonov_1971,Dyakonov_1971_2,Hirsch_1999,Murakami_2003,Sinova_2015}, etc.
} %
Spin waves (or magnons) in magnetic materials,
namely the dynamics of the ferromagnetic or antiferromagnetic order parameters,
are also elementary excitations that carry spin and heat currents \cite{Slonczewski_1989,Meier_2003,Wang_2004,Kajiwara_2010,Chumak_2015}.
Magnon spin current can be generated by the magnetic resonance under a microwave \cite{Sandweg_2011,Chumak_2012},
by the spin Seebeck effect under a temperature gradient \cite{Uchida_2010,Uchida_2010_2}, etc.

\begin{figure}[bp]
    \includegraphics[width=8.4cm]{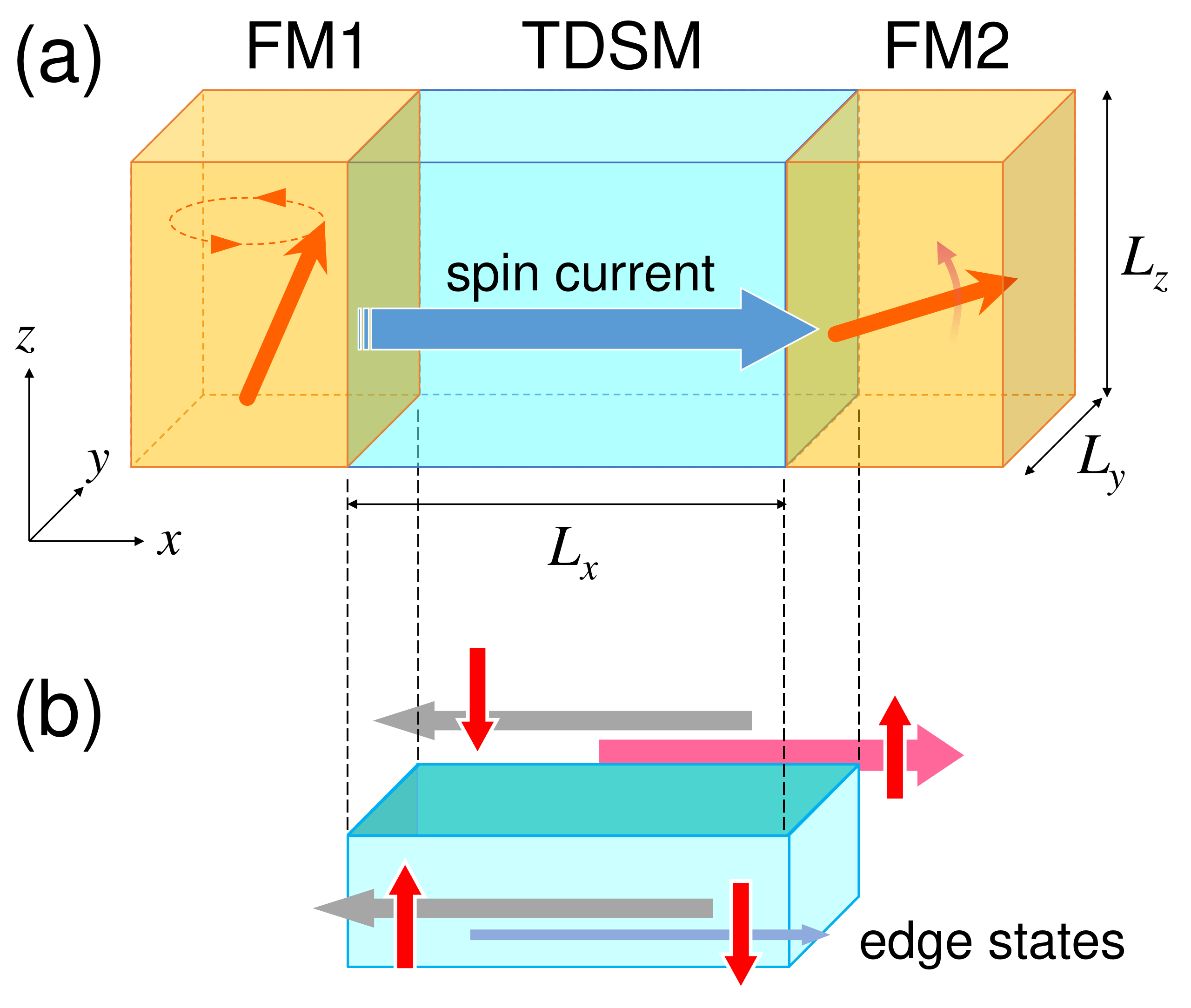}
    \caption{(a) Schematic illustration of our setup for a joint system 
    between a topological Dirac semimetal (TDSM) and ferromagnetic insulators (FMIs).
    The magnetization dynamics
    in the left-side FMI (FM1) induces a spin current flowing through the TDSM.
    We investigate the torque exerted on the right-side FMI (FM2).
    (b) Schematic illustration of the spin current carried by the helical surface states of the TDSM.
    There are four spin-polarized channels on the surface of the TDSM connecting FM1 and FM2.
    The magnetization dynamics induces population imbalance among the four edge channels, which yields a net spin current from FM1 to FM2.}
    \label{fig:schem}
\end{figure}

While those spin-current carriers are available in various materials commonly used in experiments,
\modify{
    their propagation length is inevitably limited due to dissipation.
}%
Conduction electrons are subject to scattering by phonons and \modify{disorder},
which results in the Joule heating.
Magnon spin current in insulators is considered to be advantageous in that it is free from the Joule heating \cite{Cornelissen_2015,Cornelissen_2016,Giles_2017,Brataas_2020},
whereas the Gilbert damping of spins leads to the dissipation of spin and energy.
Due to such dissipation effects,
transmission of spin current over a long range is a challenging problem.
Recent theoretical and experimental studies showed that phonons in nonmagnetic insulators
may realize long-range transport of spin current \cite{Ogawa_2015,Kikkawa_2016,Hashimoto_2017,Streib_2018,Ruckeriegel_2020,An_2020}:
since circularly-polarized transverse phonons carry angular momentum,
they can mediate spin current between magnets via magnetoelastic coupling.
However, for the efficient interconversion of magnons and phonons,
one needs fine tuning of the magnon frequency.
\modify{
    Such limitations in long-range spin transport restrict the design of highly integrated spintronics devices.
}

In the present work,
we theoretically propose a long-range transmission of spin mediated by
the surface electronic states of topological Dirac semimetals (TDSMs).
TDSM is a class of three-dimensional (3D) crystalline materials having pair(s) of Dirac nodes
in the electronic band structure in the bulk \cite{Yang_2014,Burkov_2016,Taguchi_2020}.
The TDSM phase is experimentally realized in $\mathrm{Na_3 Bi}$ \cite{Wang_2012,Liu_2014} and $\mathrm{Cd_3 As_2}$ \cite{Wang_2013,Neupane_2014,Uchida_2017,Uchida_2019,Crassee_2018}.
TDSMs have quasi-1D gapless states on the surface,
which arise as Fermi arcs connecting the Dirac points projected onto the surface Brillouin zone \cite{Gorbar_2015,Yi_2014}.
These surface states are spin helical,
i.e. spin-$\uparrow$ and spin-$\downarrow$ states propagate along the surface oppositely to each other,
and are protected by the $\mathbb{Z}_2$ topology in the bulk \cite{Yang_2015,Fang_2015}.
These features are analogous with the helical edge states of 2D quantum spin Hall insulators (QSHIs) \cite{Kane_2005,Bernevig_2006,Bernevig_2006_2},
and hence the surface states of TDSMs are robust against \modify{disorder} as long as the system preserves time-reversal symmetry \cite{Nishihaya_2019,Kobayashi_2020}.
From these features,
we can expect that the helical surface states of TDSMs are suitable for long-range spin transmission.

Indeed, in the context of 2D QSHIs,
it was theoretically proposed in numerous literatures
that the helical edge states are capable of interconversion between spin angular momentum and electric current,
based on the topological field theory, the numerical simulations,
the scattering theory, and the Floquet theory \cite{Qi_2008,Chen_2010,Mahfouzi_2010,Hattori_2013,Meng_2014,Deng_2015,Wang_2019,Araki_2020}.
The electric current and the spin torque arising from the interconversion take quantized values
irrespective of the microscopic coupling structure between the edge electrons and the spins in magnets,
which is traced back to the band topology of the 1D edge states.
The similar spin-charge interconversion is expected also on the helical surface states of TDSMs,
as long as the Fermi level is tuned in the vicinity of the Dirac points
so that the bulk transport may be negligible \cite{Misawa_2019}.
\modify{
    However, the spin-charge conversion discussed in those works
    occurs locally at the interface of a magnet and a TDSM (or a QSHI),
    and a theory for nonlocal transmission of spins with the helical edge states over a long distance,
    which is essential for device application, is not well established.
}

Based on the above background,
we here consider the transmission of spin angular momentum 
between two ferromagnetic insulators (FMIs) connected by a TDSM, as schematically shown in Fig.~\ref{fig:schem} (a).
We assume a magnetization dynamics in one of the FMIs (FM1),
and discuss how the spin current transmitted through the TDSM exerts a spin torque on the other FMI (FM2),
\modify{by constructing analytical and numerical schemes to evaluate the nonlocal spin transmission between the two FMIs separated at a distance.}
As a result, we find that the transmitted spin current takes a semi-quantized value,
which is determined only by the configuration of Dirac points in momentum space
and the frequency of magnetization dynamics.
This semi-quantization of spin current can be understood analytically as the electron transport on the helical surface states,
which comes from the imbalance of electron population among the edge channels driven by the magnetization dynamics [see Fig.~\ref{fig:schem}(b)].
Moreover, from the numerical simulation of the real-time dynamics of electrons in the whole 3D system,
we directly confirm that this semi-quantized spin transmission is robust against moderate \modify{disorder} in the bulk.
\modify{
    These results imply that the TDSM surface may serve as a promising system for highly-integrated spintronics devices with a long-range spin transmission.
}

This article is organized as follows.
In Section \ref{sec:setup},
we review the generic characteristics of TDSMs,
and give a detailed explanation about our model setup with a TDSM and FMIs shown in Fig.~\ref{fig:schem}.
In Section \ref{sec:edge-picture},
we give analytical expressions of the flow of electrons and spin on the surface,
based on the 1D scattering theory.
(The detailed calculation processes are shown in Appendices.)
In Section \ref{sec:simulation},
we show the results of our numerical simulations within the whole 3D system,
and discuss their consistency with the analytical expressions and their robustness against \modify{disorder}.
Finally, in Section \ref{sec:conclusion},
we give some experimental implications from our calculations
and conclude our discussion.
Throughout this article,
we take the natural unit $\hbar =1$.

\section{Model setup with TDSM} \label{sec:setup}
In order to demonstrate spin transmission through a TDSM,
we construct a model
that we shall use both for the analysis and for the numerical simulation,
as shown in Fig.~\ref{fig:schem}.
We first review the generic characteristics of TDSMs,
and give a detailed explanation about our model setup,
with a junction of a TDSM and two ferromagnetic insulators (FMIs),
on the basis of those characteristics.

\begin{figure}[bp]
    \includegraphics[width=7cm]{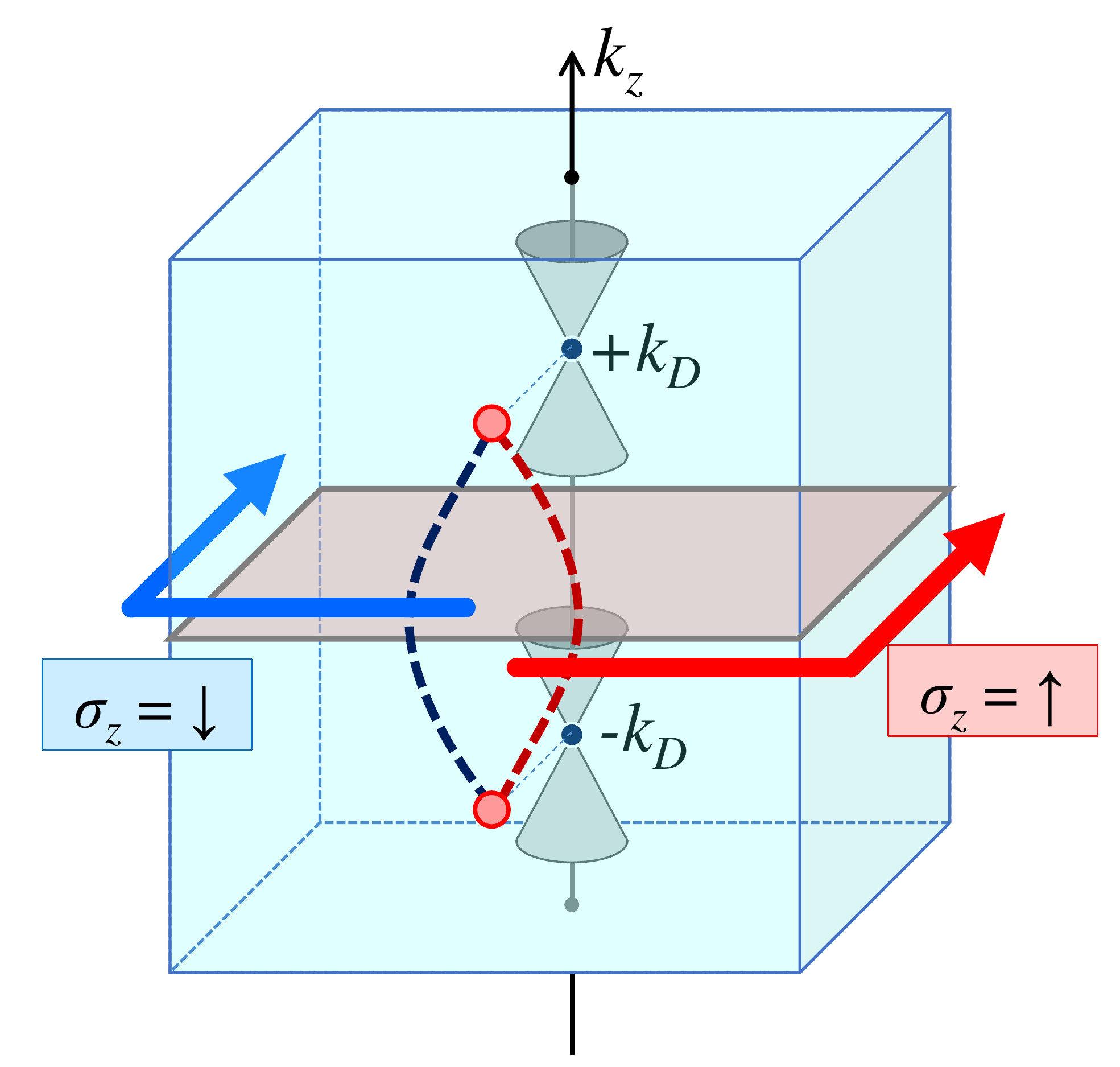}
    \caption{Schematic picture of the structure of the surface states of TDSM.
    The surface states arise as Fermi arcs connecting the Dirac points projected onto the surface Brillouin zone,
    shown as dark red and blue dashed curves.
    They form a pair of counterpropagating states,
    one with spin $\sigma_z =\uparrow$ and the other with $\sigma_z = \downarrow$.
    Taking the 2D slice in the band-inverted region ($k_z \in [-k_D,k_D]$, shown by the gray plane),
    the Hamiltonian reduces to that of 2D QSHI.}
    \label{fig:dirac-points}
\end{figure}

TDSMs are characterized by a pair of Dirac points in momentum space,
which are protected by rotational symmetry around a crystal axis \cite{Yang_2014,Yang_2015,Fang_2015}.
If we take this axis as $z$-axis, the Dirac points are located on $k_z$-axis,
which we denote as $\boldsymbol{k}_D^\pm = (0,0,\pm k_D)$.
Due to the rotational symmetry, the spin component $\sigma_z$ serves as a good quantum number around $k_z$-axis,
which means that the spin-$\uparrow$ and spin-$\downarrow$ states are degenerate around the Dirac points.
The minimal model for such a band structure at low energy is composed of four degrees of freedom,
with twofold spins and twofold orbitals \cite{Burkov_2016,Wang_2012,Wang_2013},
\begin{align}
    H(\bk) &= v(k_x \tau_x \sigma_z + k_y \tau_y) -M(k_z)\tau_z, \label{eq:model-TDSM} \\
    M(k_z) &= m_0 - m_1 k_z^2. \nonumber
\end{align}
Here the Pauli matrices $\sigma_{x,y,z}$ act on the spin subspace
and $\tau_{x,y,z}$ on the orbital subspace.
In the effective model of $\mathrm{Cd_3 As_2}$ \cite{Wang_2013}, for instance,
the basis functions with $\tau_z =+$ correspond to the $5s$-orbitals of Cd with the total angular momentum $J_z = \pm 1/2$,
while those with $\tau_z =+$ correspond to the $4p$-orbitals of As with $J_z = \pm 3/2$,
and $\sigma_z=\pm$ represents the sign of $J_z$.
The parameter $m_0$ characterizes the band inversion,
which leads to the Dirac points of $k_D = \sqrt{m_0/m_1}$ if $m_0,m_1 >0$.


Due to the band inversion from spin-orbit coupling, the system shows the intrinsic spin Hall effect.
By fixing $k_z$ in the band-inverted region $-k_D < k_z < k_D$ and considering the 2D slice,
as shown in Fig.~\ref{fig:dirac-points},
the Hamiltonian reduces to that for the 2D quantum spin Hall insulator (QSHI),
with the quantized spin Hall conductivity $\sigma_{xy}^{s\mathrm{(2D)}} = e/2\pi$.
Therefore, by multiplying the number of the 2D slices $\nu_z = 2k_D/2\pi$ per unit length in $z$-direction,
the spin Hall conductivity of the TDSM in 3D takes the ``semi-quantized'' value
\begin{align}
    \sigma_{xy}^{s\mathrm{(3D)}} = \nu_z \sigma_{xy}^{s\mathrm{(2D)}} = \frac{\modify{e} k_D}{2\pi^2} ,
\end{align}
if the Fermi level is in the vicinity of the Dirac points \cite{Burkov_2016,Taguchi_2020}.

Another consequence of the band inversion is the emergence of surface states.
On the surfaces parallel to the rotational axis ($z$-axis),
which we call the side surfaces,
there emerge spin-helical states, with the spin-$\uparrow$ and spin-$\downarrow$ modes propagating along the surface oppositely to each other \cite{Gorbar_2015}.
These surface states can be regarded as the collection of
1D helical edge states of the 2D QSHI at fixed $k_z$.
They form a pair of Fermi arcs connecting the Dirac points projected onto the surface Brillouin zone,
which are robust against disorder as long as time-reversal symmetry is preserved \cite{Kobayashi_2020}.
The contribution of these surface states to the electron transport was observed experimentally,
as the quantum oscillation under a magnetic field \cite{Uchida_2017,Nishihaya_2019,Potter_2014,Moll_2016,Zhang_2017,Zheng_2017,Zhang_2019,Lin_2019}.

In order to make use of the helical surface states for spin transmission,
here we consider the model setup as shown in Fig.~\ref{fig:schem},
with two FMIs (FM1 and FM2) attached on the side surfaces of the TDSM.
FM1 and FM2 are set apart by the distance $L_x$,
and each of them is attached to the TDSM by the length $L_y$.
We assume a situation where the magnetization of FM1 is steadily precessing around $z$-axis,
which is maintained by providing angular momentum and energy externally by microwaves, etc.
Under such a setup, we estimate the spin torque exerted on FM2,
which corresponds to the spin current transmitted from FM1 to FM2 via the TDSM,
both analytically and numerically.

\section{Transport analysis on the surface} \label{sec:edge-picture}
In this section, we treat the spin transmission through the TDSM analytically,
by focusing on the spin transport mediated by the helical surface states on the surface.
If the Fermi level is in the vicinity of the Dirac points,
the bulk transport becomes negligible and the surface transport becomes dominant.
In order to evaluate the spin transmission between two FMIs phenomenologically,
we first formulate the transmission of charge and spin at a single interface with a FMI.
By using this single-FMI picture as a building block,
we formulate the spin transmission between the two FMIs in our model setup shown in Fig.~\ref{fig:schem}.

As mentioned in the previous section,
we regard the helical surface states of the TDSM as the collection of 1D helical edge states,
which reside at every $k_z$ in the band-inverted region $(-k_D < k_z < k_D)$.
As long as the translational symmetry in $z$-direction is satisfied,
$k_z$ serves as a good quantum number,
and the contribution from 1D helical edge states at each $k_z$ can be treated separately.
Therefore, in this section,
we first consider the spin transmission by a single pair of 1D helical edge states,
and then multiply the number of 2D slices $\nu_z = 2k_D/2\pi$ per unit length in $z$-direction,
to evaluate the overall contribution from the 2D surface states.

\subsection{Charge and spin pumping by a single FMI} \label{sec:pumping}

\begin{figure}[tbp]
    \includegraphics[width=8.4cm]{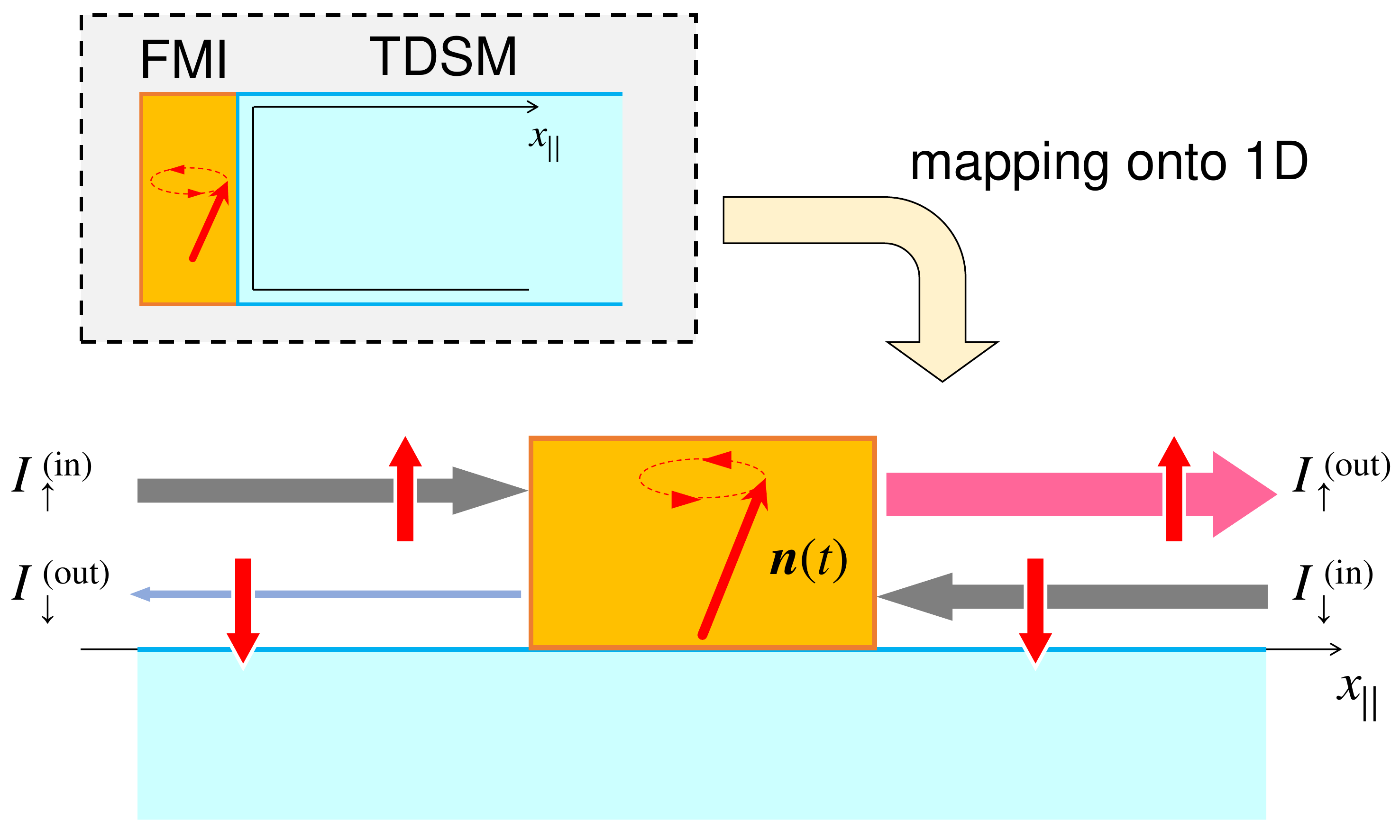}
    \caption{Schematic picture for the pumping process of the edge electrons by a single FMI.
    The pumping process is mapped to the scattering problem in the quasi-1D space $(x_\parallel)$ along the edge,
    by regarding the interface region with the precessing magnetization $\boldsymbol{n}(t)$ as the time-dependent scatterer.
    By solving the scattering problem as described in Section \ref{sec:scattering-1d},
    we see that the outgoing channel with spin-$\uparrow$ becomes more populated than that with spin-$\downarrow$.}
    \label{fig:edge-1d}
\end{figure}

In a manner similar to the theoretical treatment of spin pumping and injection by a ferromagnet \cite{Tserkovnyak_2002,Tserkovnyak_2005},
we formulate the role a FMI coupled to the helical edge as a time-dependent scatterer.
We consider the scattering process in the hypothetical 1D space along the edge,
where we denote its 1D coordinate as $x_\parallel$ (see Fig.~\ref{fig:edge-1d}).

As in the conventional Landauer--B\"{u}ttiker formalism in mesoscopic systems \cite{Landauer_1957,Buttiker_1986,Buttiker_1992,Buttiker_1994,Datta_1995},
the charge and spin currents can be derived by comparing the numbers of the incoming and outgoing electrons for the scatterer.
Since the nonmagnetic edge states are spin-helical,
there are two incoming channels and two outgoing channels:
as the incoming channels,
spin-$\uparrow$ electrons come from the left $(x_\parallel <0)$ and spin-$\downarrow$ electrons come from the right $(x_\parallel>0)$.
As the outgoing channels,
spin-$\uparrow$ electrons go out to the right $(x_\parallel >0)$
and spin-$\downarrow$ electrons go out to the left $(x_\parallel<0)$.
\modify{
    We denote the annihilation (creation) operators of an electron with energy $\epsilon$ in these channels as
    $a_{\uparrow/\downarrow}^{(\dag)}(\epsilon)$ for the incoming channels
    and $b_{\uparrow/\downarrow}^{(\dag)}(\epsilon)$ for the outgoing channels,
    respectively.
}

\modify{
    The electron distributions in these channels with energy $\epsilon$
    are given by taking the quantum average $\langle \cdot \rangle$ of the operators defined above \cite{Buttiker_1986,Buttiker_1992,Buttiker_1994},
    \begin{align}
        f^{\mathrm{(in)}}_{\uparrow / \downarrow}(\epsilon) = \langle a_{\uparrow/\downarrow}^{\dag}(\epsilon) a_{\uparrow/\downarrow}(\epsilon) \rangle, \quad
        f^{\mathrm{(out)}}_{\uparrow / \downarrow}(\epsilon) = \langle b_{\uparrow/\downarrow}^{\dag}(\epsilon) b_{\uparrow/\downarrow}(\epsilon) \rangle, 
        \label{eq:distribution}
    \end{align}
    which we use throughout this section as the main tool to evaluate the transmission of spin current.
    With these distribution functions,
}
the numbers of electrons coming into / going out of the scatterer with spin-$\uparrow/\downarrow$ per unit time
are given by
\begin{align}
    I^{\mathrm{(in/out)}}_{\uparrow/\downarrow} &= \frac{1}{2\pi} \int d\epsilon \ f^{\mathrm{(in/out)}}_{\uparrow/\downarrow}(\epsilon). \label{eq:currents}
\end{align}
By using these notations,
the charge current flowing from the left to the right is given by
\begin{align}
    I &= -e[I^{\mathrm{(in)}}_{\uparrow} - I^{\mathrm{(out)}}_{\downarrow}] 
    = -e[I^{\mathrm{(out)}}_{\uparrow} - I^{\mathrm{(in)}}_{\downarrow}]. \label{eq:charge-current}
\end{align}
The two formalisms are equivalent due to the charge conservation at the scatterer.
On the other hand, spin can be transferred between the electrons and the FMI,
and thus the net spin current flowing out of the scatterer can be nonzero.
Noting that each electron carries spin $\pm 1/2$,
the spin current pumped by the FMI, \modify{namely the net spin angular momentum flowing into and out of the FMI per unit time}, is given by
\begin{align}
    I^s &= \frac{1}{2} \left[I^{\mathrm{(out)}}_\uparrow - I^{\mathrm{(out)}}_\downarrow - I^{\mathrm{(in)}}_\uparrow +I^{\mathrm{(in)}}_\downarrow \right]. \label{eq:spin-current}
\end{align}
From these relations,
we can immediately see a simple relation between the charge and spin currents,
\begin{align}
    \frac{1}{e} I + I^s &= - I_{\uparrow}^{\mathrm{(in)}} + I_{\downarrow}^{\mathrm{(in)}}, \label{eq:current-spin}
\end{align}
where the right-hand side is determined only by the numbers of incoming particles
and is independent of the scattering process.
In particular,
if the numbers of spin-$\uparrow$ and spin-$\downarrow$ electrons
entering the scattering region are equal,
its right-hand side vanishes and reduces to the simple relation, $I^s = -I/e$.

In order to evaluate the charge current $I$ and the spin current $I^s$ separately,
we need relations between the incoming and outgoing distributions
that are determined by the scatterer.
If the magnetization $\boldsymbol{n}$ is periodically precessing as
\begin{align}
    \boldsymbol{n}(t) =(\sin\theta \cos(\Omega t +\phi) , \sin\theta \sin(\Omega t +\phi) , \cos\theta),
\end{align}
where $\Omega$ is the precession frequency and $\theta$ is the azimuthal angle,
the energy of electron is not conserved in the scattering process.
Such a time-dependent scattering problem can be solved by taking the ``rotating frame'' of spin:
by the time-dependent unitary transformation
\begin{align}
    U(t) = e^{i \Omega t \sigma_z/2}
\end{align}
on the edge electrons,
which rotates their spin by the angle $\Omega$ per unit time around $z$-axis,
the magnetization direction is fixed to $\boldsymbol{n}_0 \equiv \boldsymbol{n}(t=0)$ in this rotating frame \cite{Meng_2014}.
\modify{
    Since this transformation $U(t)$ shifts the energies of spin-$\uparrow/\downarrow$ electrons by $\pm \Omega/2$, respectively,
    the operators in the rotating frame, which we denote by $\tilde{a}_{\uparrow/\downarrow}$ and $\tilde{b}_{\uparrow/\downarrow}$, are related to those in the rest frame as
    \begin{align}
        \tilde{a}_{\uparrow}(\epsilon) = a_\uparrow(\epsilon+\tfrac{\Omega}{2}) , \quad
        \tilde{a}_{\downarrow}(\epsilon) = a_\downarrow(\epsilon-\tfrac{\Omega}{2}), \label{eq:operator-transformation} \\
        \tilde{b}_{\uparrow}(\epsilon) = b_\uparrow(\epsilon+\tfrac{\Omega}{2}) , \quad
        \tilde{b}_{\downarrow}(\epsilon) = b_\downarrow(\epsilon-\tfrac{\Omega}{2}). \nonumber
    \end{align}
}

\modify{
    The operators for the incoming and outgoing channels are related by the $S$-matrix.
    By using the $S$-matrix in the rotating frame
    \begin{align}
        \tilde{S}(\epsilon) = \begin{pmatrix}
            \tilde{r}_{\downarrow\uparrow}(\epsilon) & \tilde{t}_{\downarrow\downarrow}(\epsilon) \\
            \tilde{t}_{\uparrow\uparrow}(\epsilon) & \tilde{r}_{\uparrow\downarrow}(\epsilon)
        \end{pmatrix}
        ,
    \end{align}
    which can be obtained by solving the time-independent scattering problem with the fixed magnetization (see Appendix \ref{sec:S-matrix} for detail),
    the operators $\tilde{a}_{\uparrow/\downarrow}$ and $\tilde{b}_{\uparrow/\downarrow}$ in the rotating frame are related as \cite{Buttiker_1986,Buttiker_1992,Buttiker_1994}
    \begin{align}
        \begin{pmatrix}
            \tilde{b}_\downarrow(\epsilon) \\
            \tilde{b}_\uparrow(\epsilon)
        \end{pmatrix}
        =
        \begin{pmatrix}
            \tilde{r}_{\downarrow\uparrow}(\epsilon) & \tilde{t}_{\downarrow\downarrow}(\epsilon) \\
            \tilde{t}_{\uparrow\uparrow}(\epsilon) & \tilde{r}_{\uparrow\downarrow}(\epsilon)
        \end{pmatrix}
        \begin{pmatrix}
            \tilde{a}_\uparrow(\epsilon) \\
            \tilde{a}_\downarrow(\epsilon)
        \end{pmatrix}
        .
        \label{eq:S-rotated}
    \end{align}
    Note that the components in the $S$-matrix satisfy the reversibility relations
    \begin{align}
        |\tilde{r}_{\downarrow\uparrow}(\epsilon)|^2 = |\tilde{r}_{\uparrow\downarrow}(\epsilon)|^2 &\equiv R(\epsilon) \\
        |\tilde{t}_{\uparrow\uparrow}(\epsilon)|^2 = |\tilde{t}_{\downarrow\downarrow}(\epsilon)|^2 &\equiv T(\epsilon)
    \end{align}
    and the unitarity condition
    \begin{align}
        R(\epsilon) + T(\epsilon) =1
    \end{align}
    due to the time-independence of the scatterer in the rotating frame,
    where $R(\epsilon)$ is the reflection rate and $T(\epsilon)$ is the transmission rate.
}

\modify{
    With the $S$-matrix defined above, we are ready to evaluate the electron distributions in the outgoing channels.
    By substituting Eqs.~(\ref{eq:operator-transformation}) and (\ref{eq:S-rotated}) to Eq.~(\ref{eq:distribution}),
    we obtain the relations for the distribution functions in the rest frame as
    \begin{align}
        f^{\mathrm{(out)}}_\uparrow(\epsilon+\tfrac{\Omega}{2}) &= T(\epsilon) f^{\mathrm{(in)}}_\uparrow(\epsilon+\tfrac{\Omega}{2}) + R(\epsilon) f^{\mathrm{(in)}}_\downarrow(\epsilon-\tfrac{\Omega}{2}) \label{eq:in-out-1}\\
        f^{\mathrm{(out)}}_\downarrow(\epsilon-\tfrac{\Omega}{2}) &= R(\epsilon) f^{\mathrm{(in)}}_\uparrow(\epsilon+\tfrac{\Omega}{2}) + T(\epsilon) f^{\mathrm{(in)}}_\downarrow(\epsilon-\tfrac{\Omega}{2}). \label{eq:in-out-2}
    \end{align}
    which we shall use as the fundamental relations throughout this section to evaluate the spin transmission.
}
%
By integrating over the energy $\epsilon$ and using the unitarity relation $R(\epsilon) + T(\epsilon) =1$,
we obtain the relations between the numbers of incoming and outgoing electrons,
\begin{align}
    I^{\mathrm{(out)}}_\uparrow - I^{\mathrm{(in)}}_\uparrow &=  \int \frac{d\epsilon}{2\pi} R(\epsilon-\tfrac{\Omega}{2}) \left[ f^{\mathrm{(in)}}_\downarrow(\epsilon-\Omega) - f^{\mathrm{(in)}}_\uparrow(\epsilon) \right] \label{eq:I-in-out-up} \\
    I^{\mathrm{(out)}}_\downarrow - I^{\mathrm{(in)}}_\downarrow &=  \int \frac{d\epsilon}{2\pi} R(\epsilon+\tfrac{\Omega}{2}) \left[ f^{\mathrm{(in)}}_\uparrow(\epsilon+\Omega) - f^{\mathrm{(in)}}_\downarrow(\epsilon) \right]. \label{eq:I-in-out-down}
\end{align}


\subsection{Scattering rates and quantized pumping} \label{sec:scattering-1d}

\begin{figure}[tbp]
    \includegraphics[width=8.2cm]{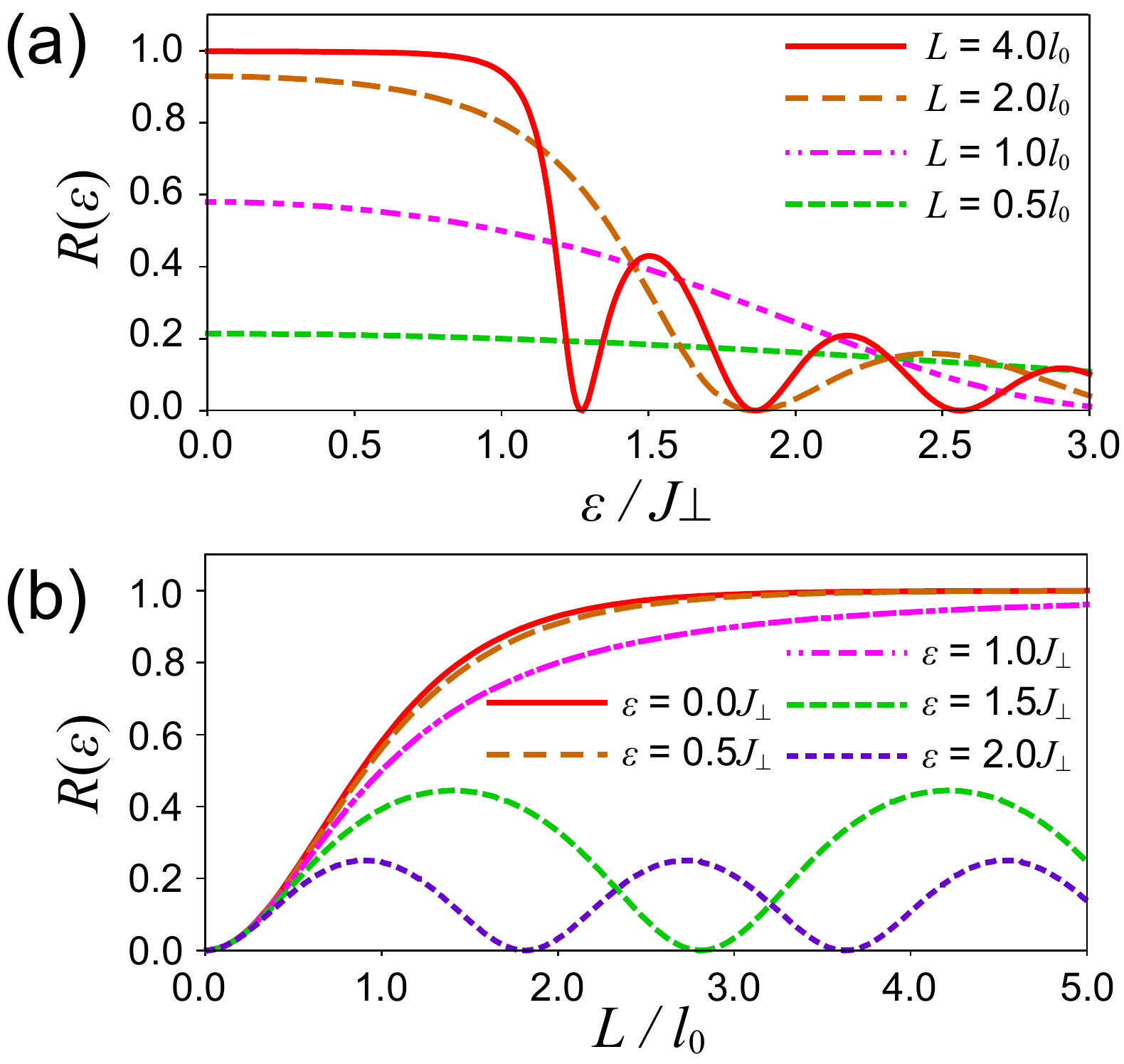}
    \caption{The reflection rate ${R}({\epsilon})$ of the magnetic region given by Eq.~(\ref{eq:Ref-rate}),
    as functions of (a) the incident energy ${\epsilon}$ and (b) the length $L$ of the magnetic region.
    $L$ is rescaled by $l_0 = v_{\mathrm{edge}}/J_\perp$,
    which is the decay length of the wave function inside the exchange gap of the magnetic region.}
    \label{fig:Reflection-rate}
\end{figure}

The scattering rates $R(\epsilon)$ and $T(\epsilon)$
are defined in the rotating frame of spin,
by a FMI with its magnetization fixed to the direction $\boldsymbol{n}_0$.
The Hamiltonian for the edge electrons coupled to this magnetization in the rotating frame
is given as
\begin{align}
    \tilde{H}(k_\parallel) &= v_{\mathrm{edge}} k_\parallel \sigma_z + J \boldsymbol{n}_0\cdot\boldsymbol{\sigma} -\tfrac{\Omega}{2}\sigma_z \\
    &=
    \begin{pmatrix}
        v_{\mathrm{edge}} k_\parallel + J\cos\theta -\tfrac{\Omega}{2} & J \sin\theta e^{-i\phi} \\
        J \sin\theta e^{i\phi} & -v_{\mathrm{edge}} k_\parallel - J\cos\theta +\tfrac{\Omega}{2}
    \end{pmatrix}
    , \nonumber
\end{align}
where we define the 1D momentum along the edge as $k_\parallel$.
We can immediately see from this matrix form that
the in-plane component of the magnetization opens an exchange gap $J_\perp \equiv |J\sin\theta|$
around zero energy in the edge spectrum,
which influences the scattering of the edge electrons by the FMI
as we shall see in the following discussions.

By evaluating the $S$-matrix in the rotating frame,
whose detailed derivation process is shown in Appendix \ref{sec:S-matrix},
the scattering rates are given as
\begin{align}
    {R}({\epsilon}) &= \frac{\sin^2 (KL)}{{\epsilon}^2/J_\perp^2 -\cos^2(KL)} \label{eq:Ref-rate} \\
    {T}({\epsilon}) &= \frac{{\epsilon}^2/J_\perp^2 -1}{{\epsilon}^2/J_\perp^2 -\cos^2(KL)}, \label{eq:Trans-rate}
\end{align}
with $K = \sqrt{\epsilon -J_\perp^2} / v_{\mathrm{edge}}$ the wave number inside the interface region coupled with the FMI.
(Note that these forms are valid for $\epsilon$ inside the exchange gap, $|\epsilon|<J_\perp$, as well,
where $K$ becomes pure imaginary and the wave function inside the interface region exponentially decays by $x_\parallel$.)
The numerical behavior of $R(\epsilon)$ is shown in Fig.~\ref{fig:Reflection-rate},
by varying the electron energy $\epsilon$ and the length of the interface region $L$.

The most important feature in the reflection rate $R(\epsilon)$ is that
it reaches unity for $|\epsilon|<J_\perp$,
which means that the electron inside the exchange gap is totally reflected,
if the length $L$ of the magnetic region is long enough.
This is because the electron wave function in the magnetic region, at energies inside the exchange gap, decays exponentially.
The decay length of the wave function at $\epsilon =0$
is given as
\begin{align}
    l_0 = [\im K_{\epsilon=0}]^{-1} = v_{\mathrm{edge}}/J_\perp,
\end{align}
and hence the tunneling through the magnetic region is fully suppressed if $L \gg l_0$.
On the other hand, if $\epsilon$ is out of the exchange gap,
the reflection rate $R(\epsilon)$ oscillates as a function of $\epsilon$
due to the formation of resonance states inside the interface region.

By using the scattering rates obtained above,
we can evaluate the charge and spin currents pumped by the FMI.
If we assume that the distributions of the incoming channels $f^{\mathrm{(in)}}_{\uparrow/\downarrow}(\epsilon)$ are in equilibrium,
with both of them filled up to the Fermi level $\epsilon_F$,
Eqs.~(\ref{eq:I-in-out-up}) and (\ref{eq:I-in-out-down}) yield the balance of incoming and outgoing electron numbers,
\begin{align}
    I^{\mathrm{(out)}}_\uparrow - I^{\mathrm{(in)}}_\uparrow = I^{\mathrm{(in)}}_\downarrow - I^{\mathrm{(out)}}_\downarrow \approx \frac{\Omega}{2\pi} R(\epsilon_F), \label{eq:I-balance-adiabatic}
\end{align}
up to the first order in $\Omega$ for slow magnetization dynamics.
\modify{
    This relation means that the number rate of outgoing electrons with spin-$\uparrow$ is raised
    and that with spin-$\downarrow$ is lowered by $R(\epsilon_F) \Omega/2\pi$ due to the magnetization dynamics,
}
as schematically shown in Fig.~\ref{fig:edge-1d}.

In particular, if the Fermi level $\epsilon_F$ is inside the exchange gap $(|\epsilon_F| < J_\perp)$,
the electrons at the Fermi level are fully reflected, i.e. $R(\epsilon_F) \approx 1$.
The right-hand side of Eq.~(\ref{eq:I-balance-adiabatic}) thus reduces to $\Omega/2\pi$,
which corresponds to one electron per a precession period $T_p = 2\pi/\Omega$.
As a consequence, the electric current pumped through the magnetic region becomes quantized as
\begin{align}
    \bar{I} = -e \frac{\Omega}{2\pi},
\end{align}
which is consistent with the previous literatures on the helical edge states of QSHI \cite{Qi_2008,Chen_2010,Mahfouzi_2010,Hattori_2013,Meng_2014,Deng_2015,Wang_2019,Araki_2020}.
The spin injection rate (per unit time) from the FMI into the edge electrons is also quantized as
\begin{align}
    \bar{I}^s = \frac{\Omega}{2\pi},
\end{align}
which satisfies the relation in Eq.~(\ref{eq:current-spin}).
The overall contribution from the 2D surface states of TDSM is given by multiplying the factor $\nu_z = 2k_D/2\pi$
to those quantized values.


\subsection{Spin transfer between two FMIs} \label{ssec:edge-two-FMI}

\begin{figure}[tbp]
    \includegraphics[width=8.4cm]{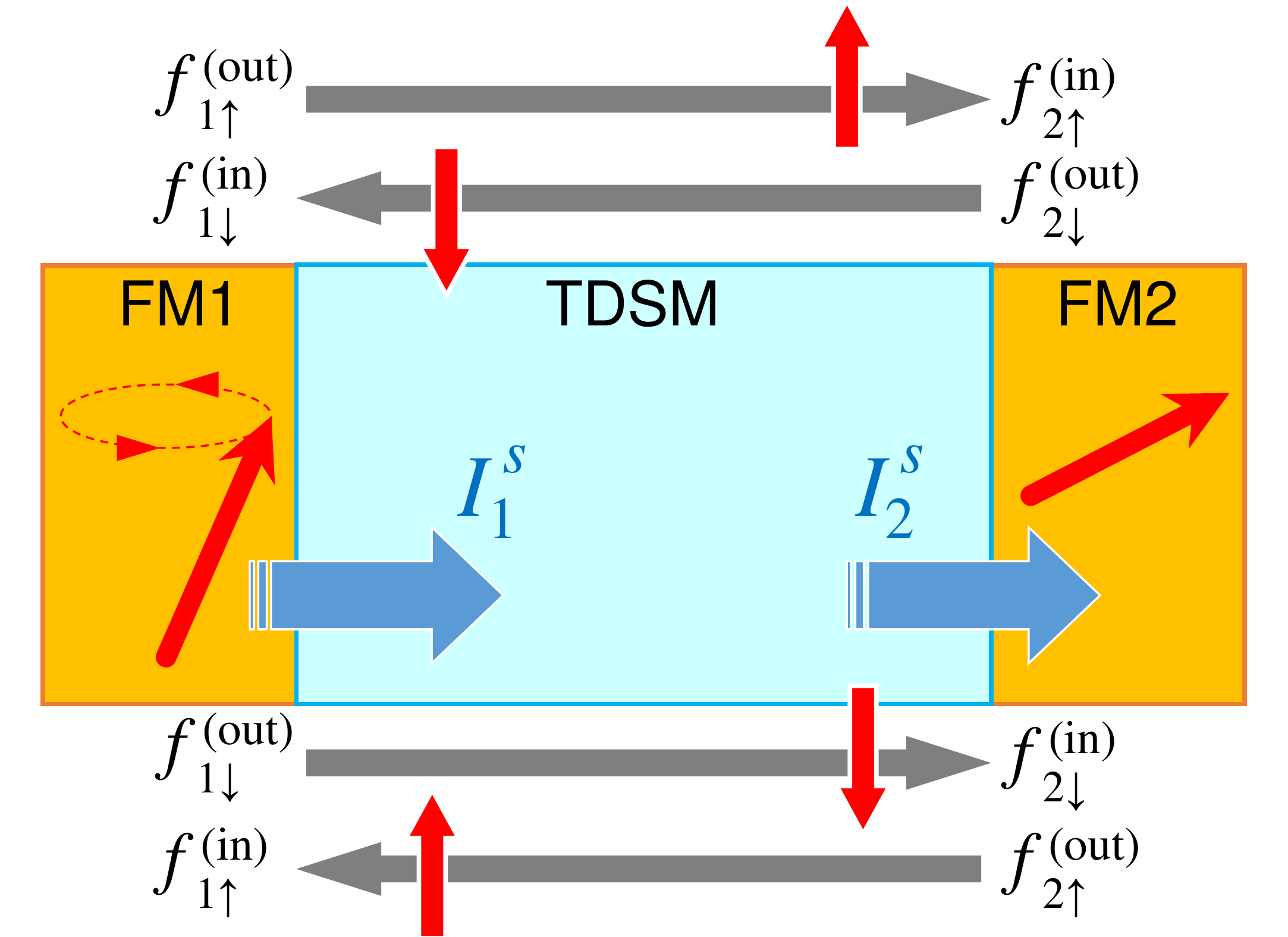}
    \caption{Schematic picture of the model setup shown in Fig.~\ref{fig:schem} sliced at a fixed $k_z$.
    On this 2D slice, FM1 and FM2 are connected by four edge channels, two with spin-$\uparrow$ and two with spin-$\downarrow$.
    We consider the electron distributions in these channels, to understand the flow of spin $I^s_1$ and $I^s_2$.}
    \label{fig:fm-qshi-fm}
\end{figure}

We now consider the model setup constructed in Section \ref{sec:setup}.
By slicing the system at fixed $k_z$ in the band-inverted region $(-k_D < k_z < k_D)$,
FM1 and FM2 are connected by four channels,
namely two pairs of counterpropagating modes with spin-$\uparrow$ and $\downarrow$,
as shown in Fig.~\ref{fig:fm-qshi-fm}.
We denote the distribution functions for the incoming/outgoing electrons
with spin $s (= \uparrow,\downarrow)$ at FM$i$ $(i=1,2)$ as $f_{is}^{\mathrm{(in/out)}}(\epsilon)$.
If the electrons on the edges propagate coherently on these channels,
$f_{1\uparrow}^{\mathrm{(out)}}(\epsilon)$ at a certain time
is equal to $f_{2\uparrow}^{\mathrm{(in)}}(\epsilon)$ after a time $T_x = L_x/v_{\mathrm{edge}}$,
and in similar manners for the other channels.
Electron propagation on these channels leads to spin transmission between FM1 and FM2.
We do not take into account the contribution from the bulk electrons here,
which is a valid approximation if the Fermi level is set in the vicinity of the Dirac points
so that the density of states in the bulk is small enough.

Now we consider the spin transmission,
with the magnetization in FM1 precessing around $z$-axis by the frequency $\Omega$
and that in FM2 fixed in $x$-direction.
Here, the transmission and reflection coefficients at FM1 are the same as those obtained in the previous subsection (with $L \rightarrow L_y$),
and those at FM2 are given by setting $\Omega=0$.
Therefore,
in a manner similar to Eqs.~(\ref{eq:in-out-1}) and (\ref{eq:in-out-2}),
the incoming and outgoing distribution functions are related as follows:
\begin{align}
    f_{1\uparrow}^{\mathrm{(out)}}(\epsilon) &= T(\epsilon-\tfrac{\Omega}{2}) f_{1\uparrow}^{\mathrm{(in)}}(\epsilon) + R(\epsilon-\tfrac{\Omega}{2}) f_{1\downarrow}^{\mathrm{(in)}}(\epsilon-\Omega) \label{eq:f1u}\\
    f_{1\downarrow}^{\mathrm{(out)}}(\epsilon) &= T(\epsilon+\tfrac{\Omega}{2}) f_{1\downarrow}^{\mathrm{(in)}}(\epsilon) + R(\epsilon+\tfrac{\Omega}{2}) f_{1\uparrow}^{\mathrm{(in)}}(\epsilon+\Omega) \label{eq:f1d}\\
    f_{2\uparrow}^{\mathrm{(out)}}(\epsilon) &= T(\epsilon) f_{2\uparrow}^{\mathrm{(in)}}(\epsilon) + R(\epsilon) f_{2\downarrow}^{\mathrm{(in)}}(\epsilon) \label{eq:f2u}\\
    f_{2\downarrow}^{\mathrm{(out)}}(\epsilon) &= T(\epsilon) f_{2\downarrow}^{\mathrm{(in)}}(\epsilon) + R(\epsilon) f_{2\uparrow}^{\mathrm{(in)}}(\epsilon). \label{eq:f2d}
\end{align}
For simplicity of discussion,
we assume here that electron transmission and reflection at each magnetic region occur instantaneously,
which is satisfied if $L_y \ll v_{\mathrm{edge}} T_p$.

\begin{figure}[tbp]
    \includegraphics[width=8.4cm]{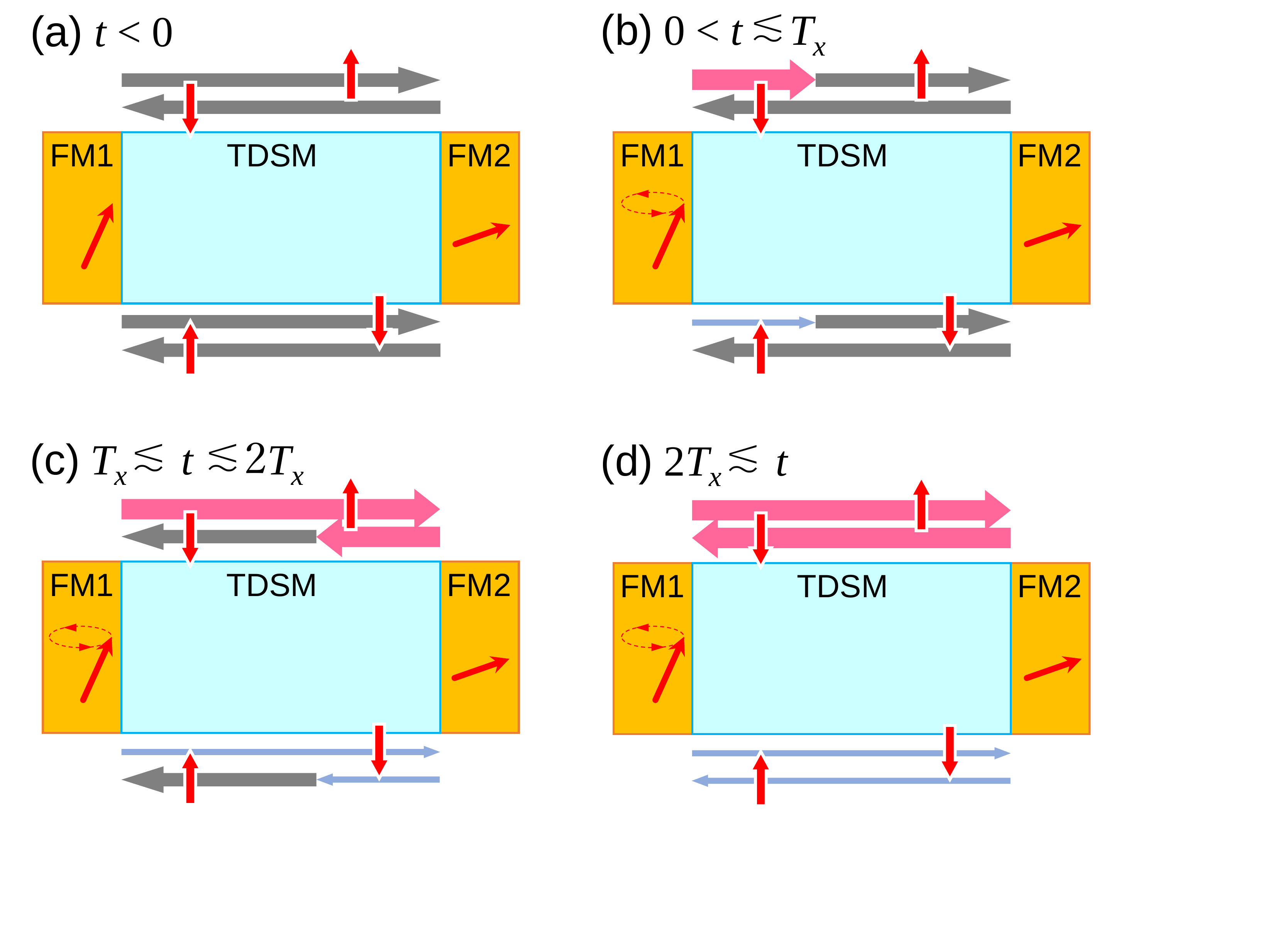}
    \caption{Schematic pictures of the electron population of edge channels,
    after switching on the magnetization dynamics in FM1 at time $t=0$.
    In a similar manner with Fig.~\ref{fig:edge-1d},
    gray arrows indicate edge channels in the equilibrium distribution,
    red thick arrows indicate channles more populated than the equilibrium distribution,
    and blue thin arrows indicate less populated channels.}
    \label{fig:edge-evolution}
\end{figure}

Based on the above relations,
we evaluate the flow of charge and spin between FM1 and FM2,
driven by the magnetization dynamics in FM1.
As the initial condition, we start with the system in equilibrium,
where all the edge channels are in the equilibrium distribution $f_0(\epsilon)$
filled up to the Fermi level $\epsilon_F$ the Fermi level.
We then switch on the magnetization dynamics in FM1 at time $t=0$ adiabatically,
so that the switch-on process may not cause any significant disturbance in the electron distributions,
and estimate the transient behavior of the edge electrons after the switch-on by considering the following steps (a)-(d).
(The schematic pictures corresponding to these steps are shown in Fig.~\ref{fig:edge-evolution}.)

\noindent \textbf{(a)}
Before the magnetization dynamics is switched on $(t<0)$,
all the edge channels are in the equilibrium distribution $f_0(\epsilon)$.

\noindent\textbf{(b)}
Soon after the switch-on, for time $0 \lesssim t \lesssim T_x$,
the channels going out from FM1 $(f_{1\uparrow/\downarrow}^{\mathrm{(out)}})$ are modulated by the magnetization dynamics,
whereas the incoming channels $(f_{1\uparrow/\downarrow}^{\mathrm{(in)}})$ are still in equilibrium distributions.
Therefore, $f_{1\uparrow}^{\mathrm{(out)}}$ and $f_{1\downarrow}^{\mathrm{(out)}}$ are given as
\begin{align}
    f_{1\uparrow}^{\mathrm{(out)}}(\epsilon) &= T(\epsilon-\tfrac{\Omega}{2}) f_0(\epsilon) + R(\epsilon-\tfrac{\Omega}{2})  f_0(\epsilon-\Omega) \label{eq:f1u-0} \\
    f_{1\downarrow}^{\mathrm{(out)}}(\epsilon) &= T(\epsilon+\tfrac{\Omega}{2})f_0(\epsilon) + R(\epsilon+\tfrac{\Omega}{2})  f_0(\epsilon+\Omega). \label{eq:f1d-0}
\end{align}
In particular,
if the Fermi level $\epsilon_F$ is inside the exchange gap
and the magnetization dynamics is adiabatic $(\Omega \ll J_\perp)$,
we can apply the same discussion as in Eq.~(\ref{eq:I-balance-adiabatic}) in the previous subsection.
\modify{
    While the number rates of the incoming electrons per unit time $I_{1\uparrow}^{\mathrm{(in)}}$ and $I_{1\downarrow}^{\mathrm{(in)}}$ are equal,
    the number rate of the outgoing electrons $I_{1\uparrow}^{\mathrm{(out)}}$ gets raised,
    and $I_{1\downarrow}^{\mathrm{(out)}}$ gets lowered by $\Omega/2\pi$,
    due to the magnetization dynamics in FM1 (see Fig.~\ref{fig:edge-evolution}(b)).
}
Therefore, both the electric current passing through the magnetic region
and the spin current pumped from FM1 reach the quantized values,
\begin{align}
    \bar{I}_1 = -e\frac{\Omega}{2\pi}, \quad
    \bar{I}_1^s = \frac{1}{2} \times \frac{\Omega}{2\pi} + \frac{-1}{2} \times \frac{-\Omega}{2\pi} = \frac{\Omega}{2\pi},
\end{align}
per a single 2D slice at $k_z$.

\noindent\textbf{(c)}
For time $T_x \lesssim t \lesssim 2T_x$,
the electrons going out from FM1 at the step (b) reach FM2,
and are reflected or transmitted by FM2.
As a consequence,
the outgoing distributions for FM1 given by Eqs.~(\ref{eq:f1u-0}) and (\ref{eq:f1d-0}) serve as the incoming distributions for FM2,
$f_{2\uparrow}^{\mathrm{(in)}}(\epsilon) =f_{1\uparrow}^{\mathrm{(out)}}(\epsilon)$ and $f_{2\downarrow}^{\mathrm{(in)}}(\epsilon) = f_{1\downarrow}^{\mathrm{(in)}}(\epsilon)$.
By comparing them with the outgoing distributions $f_{2\uparrow}^{\mathrm{(out)}},f_{2\downarrow}^{\mathrm{(out)}}$
on the basis of the scattering theory in 1D,
we see that the spin angular momentum per unit time
\begin{align}
    I_2^s &= \frac{1/2}{2\pi} \int d\epsilon \left[f^{\mathrm{(in)}}_{2\uparrow}(\epsilon) - f^{\mathrm{(in)}}_{2\downarrow}(\epsilon) - f^{\mathrm{(out)}}_{2\uparrow}(\epsilon) +f^{\mathrm{(out)}}_{2\downarrow}(\epsilon) \right]
\end{align}
is transferred from the edge electrons to FM2.
By substituting Eqs.~(\ref{eq:f2u})-(\ref{eq:f1d-0}),
this spin current reads
\begin{align}
    I_2^s
    &= \frac{1}{2\pi} \int d\epsilon \ R(\epsilon) \Bigl[ R(\epsilon-\tfrac{\Omega}{2}) \left( f_0(\epsilon-\Omega) - f_0(\epsilon)\right) \nonumber \\
    & \hspace{2cm} - R(\epsilon+\tfrac{\Omega}{2}) \left( f_0(\epsilon+\Omega) -f_0(\epsilon)\right) \Bigr], \label{eq:I-s2}
\end{align}
which is the general form applicable to arbitrary precession frequency $\Omega$
and equilibrium Fermi energy $\epsilon_F$.


In particular, if the precession frequency $\Omega$ in FM1 and the Fermi level $\epsilon_F$ are inside the exchange gap $J_\perp$ (i.e. $\Omega, \epsilon_F \ll J_\perp$),
we can again derive the quantized pumping.
\modify{
    If the interface region is sufficiently long (i.e. $L_y \gg l_0$),
    the incoming electrons in the vicinity of $\epsilon_F$ are fully reflected at FM2.
    The incoming electron with spin-$\uparrow$,
    whose number rate per unit time is raised by $\Omega/2\pi$,
    flips its spin on the reflection process at FM2
    and injects spin 1 to FM2 for each electron,
    whereas that with spin-$\downarrow$ is lowered by $\Omega/2\pi$ and injects spin $-1$ to FM2
    (see Fig.~\ref{fig:edge-evolution}(c)).
    Therefore, the spin current $I^s_2$ transmitted to FM2 reaches the universal value
    \begin{align}
        \bar{I}_2^s &= 1\times \frac{\Omega}{2\pi} -1 \times \frac{-\Omega}{2\pi} =  \frac{\Omega}{\pi},
    \end{align}
}
which means that spin 2 is injected into FM2 during a precession period $T_p$
per a single 2D slice at $k_z$.
For the 3D TDSM, the injected spin current (per unit length in $z$-direction) takes the semi-quantized value
\begin{align}
    \bar{j}_2^s = \nu_z \bar{I}_2^s = \frac{k_D \Omega}{\pi^2}. \label{eq:j-TDSM}
\end{align}
This (semi-)quantization of spin current is one of the main results in our analysis,
which is determined only by the number of helical channels on the surface
and is independent of the microscopic structures in the bulk.
%
Note that this relation is satisfied only if the electrons at the Fermi surface are fully reflected by FM2.
If $\Omega$ or $\epsilon_F$ is out of the exchange gap,
or once magnetization dynamics is driven in FM2,
some electrons are transmitted through the magnetic region of FM2,
and $I^s_2$ becomes not exactly twice of $I^s_1$.

\noindent\textbf{(d)}
For time $2T_x \lesssim t$,
the electrons reflected at FM2 at the step (c) reaches FM1.
At this step, all the edge channels are no longer in equilibrium distribution.
\modify{
    Moreover, once the magnetization in FM2 acquires dynamics,
    the incoming electrons at FM2 are no longer fully reflected.
}
Therefore, the spin currents $I^s_1$ and $I^s_2$ deviate from the (semi-)quantized values $\bar{I}^s_1$ and $\bar{I}^s_2$ at this stage.
In order to make use of the (semi-)quantization of spin current,
the magnetization dynamics in FM1 should be in a pulse shorter than the time scale $T_x$.

The important feature seen from the analysis above is that
the value of the spin current transmitted by the surface states of TDSM
(during the steps (b) and (c)) is universal,
which is determined only by the location of the Dirac points $k_D$ and the precession frequency $\Omega$,
as given by Eq.~(\ref{eq:j-TDSM}).
It only requires the existence of a sizable exchange gap in the helical surface states induced by the FMIs,
and is insensitive to the microscopic structure and value of the exchange coupling.
Moreover,
the transmitted spin current is independent of the system size $L_{x,y}$,
since only a single pair of helical edge states contribute to the spin transmission
for each 2D slice at $k_z$.
This analytical estimation is valid as long as the surface states are robustly present against \modify{disorder},
which shall be checked by the numerical simulation in the next section.

\section{Numerical simulation on lattice} \label{sec:simulation}
In this section, we present our numerical simulation of the spin transmission process via a TDSM,
which is performed with the 3D lattice model of a TDSM.
For the numerical simulation,
we use the model constructed in Section \ref{sec:setup},
with two FMIs (FM1/2) connected by a TDSM.
By following the real-time evolutions
of the wave function of all the electrons in the TDSM
and of the magnetization in FM2,
we evaluate the flow of spin driven by the magnetization dynamics in FM1.
As a result, we find that the transmitted spin current reaches the semi-quantized value
at the early stage after switching on the magnetization dynamics.
This semi-quantization behavior of the spin current agrees with the surface transport picture employed in the previous section,
which implies that the spin transmission in the TDSM is dominated by the surface states.
Moreover, we observe that this semi-quantized spin transport is robust under disorder even at a long range.

\subsection{Model}
For the numerical simulation,
we use a lattice model of a TDSM \cite{Wang_2012,Wang_2013}.
On a hypothetical cubic lattice with the lattice spacing $a$,
the tight-binding Hamiltonian
\begin{align}
    H_{\mathrm{TDSM}}(\bk) &= u \left[ \sin(ak_x) \tau_x\sigma_z + \sin(ak_y)\tau_y \right] -M(\bk)\tau_z, \nonumber\\
    M(\bk) &= r_0 - r_1 \sum_{i=x,y,z} \left[1-\cos(ak_i)\right] \label{eq:lattice-model}
\end{align}
reproduces the low-energy effective model in Eq.~(\ref{eq:model-TDSM}) around $\bk =0$,
with the correspondence of parameters $v = au, \ m_0 = r_0, \ m_1 = a^2 r_1/2$.
This lattice Hamiltonian gives a pair of Dirac points located at $\bk_D^\pm = (0,0,\pm k_D)$,
with $k_D = a^{-1} \arccos(1-r_0/r_1)$.
Throughout our calculation, we fix the parameters $r_0 = r_1 = u$,
which gives $k_D = \pi/2a$.

In real space, the Hamiltonian becomes
\begin{align}
    \mathcal{H}_{\mathrm{TDSM}} &= \sum_{\br} \sum_{i=0,x,y,z} \left[ c_{\br}^\dag D_i c_{\br + \boldsymbol{a}_i} + \mathrm{H.c.} \right]
\end{align}
in the operator formalism,
where $c_{\br}^{(\dag)}$ is the four-component annihilation (creation) operator at the lattice cite $\br$.
The vectors $\boldsymbol{a}_{i=0,x,y,z}$ are defined as
$\boldsymbol{a}_0 =0$ and $\boldsymbol{a}_{x,y,z} = a \boldsymbol{e}_{x,y,z}$,
with the Cartesian unit vectors $\boldsymbol{e}_{x,y,z}$,
and the matrices $D_{i=0,x,y,z}$ are defined as
\begin{align}
    D_0 = -\frac{r_0 - 3r_1}{2} \tau_z, &\quad
    D_z = -\frac{r_1}{2} \tau_z, \\
    D_x = \frac{iu}{2} \tau_x\sigma_z -\frac{r_1}{2} \tau_z, & \quad
    D_y = \frac{iu}{2} \tau_y -\frac{r_1}{2} \tau_z. \nonumber
\end{align}
In order to simulate the setup shown in Fig.~\ref{fig:schem}(a),
we here take open-boundary conditions in $x$- and $y$-directions,
with the size represented by $L_x$ and $L_y$.
For $z$-direction, we take a periodic boundary condition, with the size $L_z$.
The number of sites $N_{x,y,z}$ in each direction is related to the system size by $L_{x,y,z} = a N_{x,y,z}$.

The magnetizations in the FMIs are defined as macrospins,
with their directions denoted by the unit vectors
\begin{align}
    \boldsymbol{n}_{i} =(\sin \theta_i \cos\phi_i , \sin\theta_i \sin\phi_i, \cos\theta_i). \quad(i=1,2)
\end{align}
We investigate their dynamics by solving the Landau--Lifshitz--Gilbert (LLG) equation,
as we discuss in detail in the next subsection,
and hence we do not implement their dynamical properties in the Hamiltonian of the TDSM.
We require that the magnetizations $\boldsymbol{n}_{1,2}$ are coupled to the electron spins in the TDSM at the boundaries $x = 0$ and $x =L_x$, respectively.
The coupling is described by the Hamiltonian
\begin{align}
    \mathcal{H}_{\mathrm{exc}} &= J_{\mathrm{exc}} \sum_{\br}^{x=0} c_{\br}^\dag (\boldsymbol{n}_1 \cdot \boldsymbol{\Sigma}) c_{\br} + J_{\mathrm{exc}} \sum_{\br}^{x=L_x} c_{\br}^\dag (\boldsymbol{n}_2 \cdot \boldsymbol{\Sigma}) c_{\br},
\end{align}
with the phenomenological coupling constant $J_{\mathrm{exc}}$.
The matrix $\boldsymbol{\Sigma}$ characterizes how the exchange coupling depends on the atomic orbital ($s$ or $p$) that each electron in the TDSM belongs to \cite{Ominato_2019}.
Here we define it as $\boldsymbol{\Sigma} = (1+\tau_z) \boldsymbol{\sigma}$,
so that the structure of the exchange coupling shall be invariant under a $C_4$ rotation around $z$-axis.
By incorporating this coupling term,
$\mathcal{H} = \mathcal{H}_{\mathrm{TDSM}} + \mathcal{H}_{\mathrm{exc}}$
is the full Hamiltonian for the electrons in the TDSM.

\subsection{Simulation method}
Based on the lattice model defined above,
we perform a numerical simulation of the dynamics of the electrons and the magnetization.
The aim of this simulation is to evaluate the influence of the magnetization dynamics in FM1 $\boldsymbol{n}_1(t)$
on the magnetization dynamics in FM2 $\boldsymbol{n}_2(t)$,
which characterizes the spin current transmitted via the TDSM.
In order to evaluate them, we simultaneously solve the time-dependent Schr\"{o}dinger equation
\begin{align}
    i \partial_t |\Psi(t)\rangle &= \mathcal{H}(t) |\Psi(t)\rangle \label{eq:Schrodinger-TDSM}
\end{align}
for the many-body wave function $|\Psi(t)\rangle$ for all the electrons in the TDSM \cite{Suzuki_1994,Nakanishi_1997},
and the LLG equation
\begin{align}
    \dot{\boldsymbol{n}}_2(t) &= -\gamma \boldsymbol{B}_{\mathrm{eff}}(t) \times \boldsymbol{n}_2 + \alpha \boldsymbol{n}_2 \times \dot{\boldsymbol{n}}_2 \label{eq:LLG-FM2}
\end{align}
for the magnetization in FM2 $\boldsymbol{n}_2(t)$,
with $\gamma$ the gyromagnetic ratio and $\alpha$ the Gilbert damping constant.
We introduce the dynamics of $\boldsymbol{n}_1(t)$ as the input,
and do not evaluate the feedback on $\boldsymbol{n}_1(t)$ from the electron dynamics.
The Hamiltonian $\mathcal{H}(t)$ for the electrons depends on $\boldsymbol{n}_2(t)$,
and the effective magnetic field $\boldsymbol{B}_{\mathrm{eff}}(t)$ for the FM2 depends on $| \Psi(t)\rangle$ via the exchange coupling.
In particular, if we define the number and magnitude of spins in FM2 as $N_s$ and $S$,
the effective magnetic field $\boldsymbol{B}_{\mathrm{eff}}(t)$ for each spin is given by
\begin{align}
    \gamma \boldsymbol{B}_{\mathrm{eff}}(t) &= -\frac{1}{N_s S} \left\langle \frac{\partial \mathcal{H}}{\partial \boldsymbol{n}_2} \right\rangle(t) \\
    &= -\frac{J_{\mathrm{exc}}}{N_s S} \sum_{\br}^{x=L_x} \langle  c_{\br}^\dag \boldsymbol{\Sigma} c_{\br} \rangle (t), \nonumber
\end{align}
where $\langle \mathcal{O} \rangle(t)$ denotes the expectation value of the operator $\mathcal{O}$
evaluated with the many-body wave function $| \Psi(t)\rangle$.
Equations (\ref{eq:Schrodinger-TDSM}) and (\ref{eq:LLG-FM2}) are thus correlated,
from which we can evaluate the spin current transmission via the TDSM.

As the initial condition for $t<0$,
we set $\boldsymbol{n}_1(t<0) = \boldsymbol{n}_2(t<0) = \boldsymbol{e}_x$,
and take $|\Psi(t<0)\rangle$ as the Slater determinant of the occupied states in equilibrium,
where all the eigenstates in the TDSM below the Fermi energy $\epsilon_F =0$ are occupied.
At time $t=0$, we switch on the in-plane magnetization dynamics in FM1
\begin{align}
    \boldsymbol{n}_1(t) = (\cos\Omega t, \sin\Omega t, 0),
\end{align}
with the precession periodicity $T_p = 2\pi/\Omega$,
and solve the time-dependent equations (\ref{eq:Schrodinger-TDSM}) and (\ref{eq:LLG-FM2}) simultaneously.
In order to evaluate the effect of the transmitted spin current exclusively,
we neglect the Gilbert damping $\alpha$ and solve Eq.~(\ref{eq:LLG-FM2}) solely with $\boldsymbol{B}_{\mathrm{eff}}$ from the exchange coupling.
We suppose that the spins in FM2 are residing on the lattice sites at the interface $x=L_x$,
which yields $N_s = N_y N_z$,
and fix $S=1$ for the simplicity of calculation.
Throughout our simulations,
we fix $N_y = 28$ and $N_z = 16$.

\subsection{Spin current vs spin torque}
Before showing our simulation results,
we discuss how the spin torque on FM2 calculated from the simulation
is related to the spin current flowing into FM2,
to compare the simulation results with the analytical estimations given in the previous section.
\modify{
    From the torque $\dot{\boldsymbol{n}}_2(t) = -\gamma \boldsymbol{B}_{\mathrm{eff}} \times \boldsymbol{n}_2$ on FM2,
    we extract the dampinglike component
    \begin{align}
        \boldsymbol{\tau}_{\mathrm{DL}}(t) &= (\dot{\boldsymbol{n}}_2 \cdot \boldsymbol{e}_{2\theta})\boldsymbol{e}_{2\theta}(t) 
        = \frac{-\dot{n}_{2z}(t)}{\sqrt{1-n_{2z}^2}}\boldsymbol{e}_{2\theta}(t) , \label{eq:damp-torque}
    \end{align}
    where the unit vector $\boldsymbol{e}_{2\theta}(t)$ is defined by
    \begin{align}
        \boldsymbol{e}_{2\theta}(t) &= \boldsymbol{n}_2 \times \frac{\boldsymbol{n}_2 \times \boldsymbol{e}_z }{|\boldsymbol{n}_2 \times \boldsymbol{e}_z|} \\
        &= (\cos\theta_2 \cos\phi_2, \cos\theta_2 \sin\phi_2, -\sin\theta_2). \nonumber
    \end{align}
    The dampinglike torque tilts the magnetization toward $z$-axis,
    which originates from the spin angular momentum injected into the magnet.
}

\modify{
    We need to check if the surface-mediated spin current estimated in the previous section is the main contribution to the dampinglike torque $\tau_{\mathrm{DL}}(t)$.
    From the discussion in the previous section,
    the net spin current flowing into FM2 reaches the semi-quantized value
    \begin{align}
        \bar{I}^{s\mathrm{(tot)}}_2 = L_z \bar{j}^s_2 = \frac{L_z k_D \Omega}{\pi^2}
    \end{align}
    in our lattice system,
    if the spin transport is dominated by the helical surface states.
    On the normalized magnetization $\boldsymbol{n}_2(t)$ in FM2,
    this spin current may exert a spin-transfer torque \cite{Slonczewski_1989}
    \begin{align}
        \bar{\boldsymbol{\tau}}_{\mathrm{stt}}(t) &= \frac{\gamma}{M^{\mathrm{(tot)}}} \boldsymbol{n}_2 \times \left( \boldsymbol{n}_2 \times \bar{I}^{s\mathrm{(tot)}}_2 \boldsymbol{e}_z \right) \\
        &= \frac{\sqrt{1-n_{2z}^2}}{N_s S} \bar{I}^{s\mathrm{(tot)}}_2 \boldsymbol{e}_{2\theta}(t) \ \left( \equiv \bar{\tau}_{\mathrm{stt}} \boldsymbol{e}_{2\theta} \right), \nonumber
    \end{align}
    where $M^{\mathrm{(tot)}} = \gamma N_s S$ denotes the net magnetic moment in FM2.
}
%

Therefore,
\modify{
    in order to check if the surface transport picture is valid,
    we compare the dampinglike torque $\tau_{\mathrm{DL}}(t)$ with this surface-mediated spin-transfer torque $\bar{\tau}_{\mathrm{stt}}(t)$,
    by evaluating their ratio $\rho(t) \equiv \tau_{\mathrm{DL}}(t) / \bar{\tau}_{\mathrm{stt}}(t)$.
    By using the particular settings $k_D = \pi/2a$, $N_s = N_y N_z$, and $S=1$ employed in our simulation,
    this ratio can be derived from the time evolution of $\boldsymbol{n}_2(t)$,
    \begin{align}
        \rho(t) \equiv \frac{\tau_{\mathrm{DL}}(t)}{\bar{\tau}_{\mathrm{stt}}(t)} = N_y T_p\frac{-\dot{n}_{2z}(t)}{1-n_{2z}^2},
    \end{align}
    which we shall plot in the following figures.
    If this ratio $\rho(t)$ reaches unity,
    we can understand that the spin transmission in the TDSM is dominated by its surface states.
}

\subsection{Results and discussions}

\begin{figure}[tbp]
    \begin{center}
      \includegraphics[width=9cm,clip]{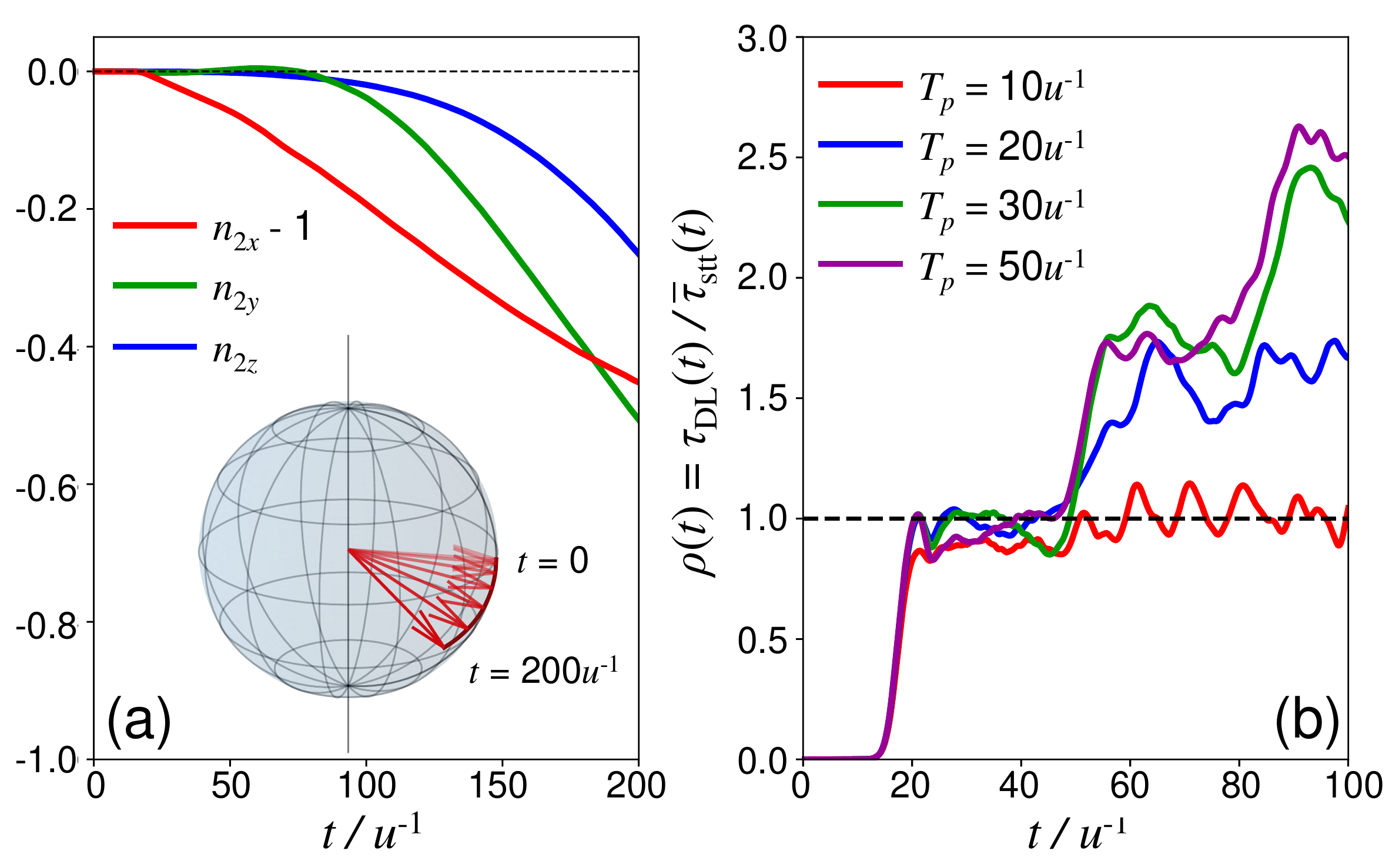}
    \end{center}
  \caption{
  (a) Time evolution of the components of $\boldsymbol{n}_2(t)$ for $J_{\mathrm{exc}} = 0.5u$, $T_p = 20u^{-1}$, and $L_x = 16a$.
  The inset shows the time evolution of $\boldsymbol{n}_2(t)$ on the Bloch sphere.
  (b) Time evolution of \modify{the ratio $\rho(t)$ of the dampinglike torque $\tau_{\mathrm{DL}}(t)$ in comparison with $\bar{\tau}_{\mathrm{stt}}(t)$ from the surface transport picture},
  with several values of $T_p$
  for $J_{\mathrm{exc}} = 0.5u$ and $L_x = 16a$.
  }
  \label{fig:depT}
\end{figure}
  
We now show our results of the numerical simulations.
First, we demonstrate a typical time-evolution behavior of  $\boldsymbol{n}_2(t)$ in Fig.~\ref{fig:depT}(a).
We here take the parameters $J_{\mathrm{exc}} = 0.5u, \ L_x = 16a$, and $T_p = 20 u^{-1}$.
As mentioned above, $\boldsymbol{n}_2$ is fixed to $x$-direction as the initial condition $(t<0)$.
After the magnetization dynamics $\boldsymbol{n}_1(t)$ in FM1 is switched on at $t=0$,
the magnetization $\boldsymbol{n}_2(t)$ in FM2 also deviates from its initial direction,
which implies that spin angular momentum is transmitted from FM1 to FM2 via the TDSM.
We can see that the out-of-plane component $n_{2z}$ evolves first at the early stage of the magnetization dynamics,
which can be considered as the effect of the dampinglike torque from the transmitted spin current.

In order to see the nature of the transmitted spin current in more detail,
we plot in Fig.~\ref{fig:depT}(b)
\modify{
    the time evolution of $\rho(t)$,
    namely the ratio of the dampinglike torque $\tau_{\mathrm{DL}}(t)$ from this simulation
    in comparison with the spin-transfer torque $\bar{\tau}_{\mathrm{stt}}(t)$ from the surface transport picture,
}
with several values of $T_p$.
We can see that, for any value of $T_p$ in these calculations,
\modify{
    $\rho(t)$ reaches unity
}
at the time $t \sim 20 u^{-1}$,
which implies that the spin current is dominated by the helical surface states of the TDSM,
as predicted in Section \ref{ssec:edge-two-FMI}.

The time evolution of the dampinglike torque $\tau_{\mathrm{DL}}^{\mathrm{(tot)}}(t)$ can be associated with the transient steps (b)-(d) described in Section \ref{ssec:edge-two-FMI} (or Fig.~\ref{fig:edge-evolution}) as follows.
Its zero-value plateau for $t \lesssim u^{-1}$
can be regarded as step (b),
with the spin signal from FM1 propagating toward FM2.
The semi-quantized plateau for $20u^{-1} \lesssim t \lesssim 40u^{-1}$ corresponds to step (c),
where the signal from FM1 is reflected by FM2 and is injecting spin angular momentum into FM2.
For $40 u^{-1} \lesssim t$,
\modify{the injected spin current}
deviates from the semi-quantized value.
This behavior can be associated with step (d),
where the signal reflected by FM2 returns to FM1 and gradually enters FM2 again,
enhancing the spin injection into FM2.



\begin{figure}[tbp]
    \begin{center}
      \includegraphics[width=9cm,clip]{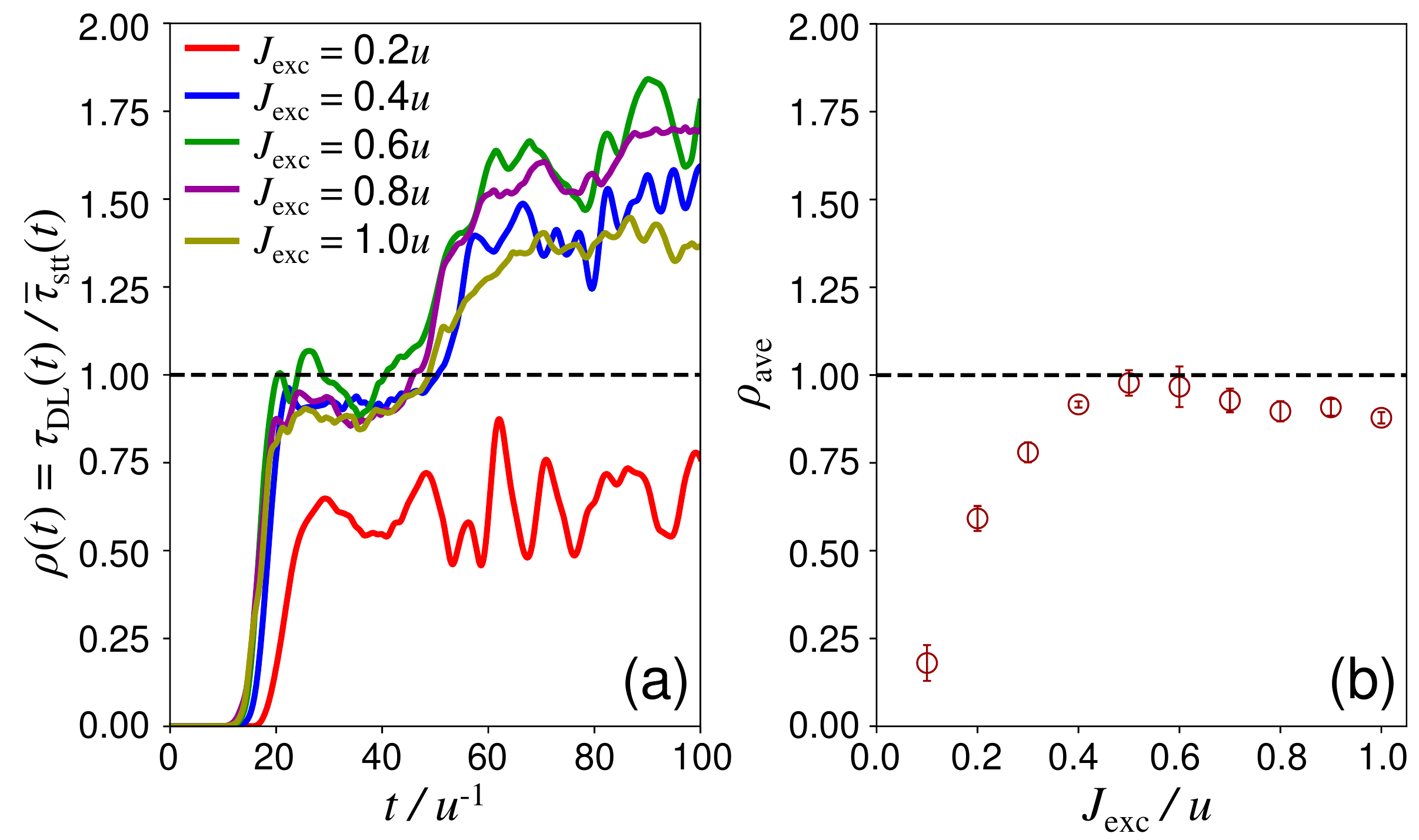}
    \end{center}
    \caption{(a) Time evolution of \modify{the ratio $\rho(t)$ of the dampinglike torque $\tau_{\mathrm{DL}}(t)$ in comparison with $\bar{\tau}_{\mathrm{stt}}(t)$ from the surface transport picture},
    with several values of $J_{\mathrm{exc}}$
    for $T_p = 20u^{-1}$ and $L_x = 16a$.
    (b) The time-averaged \modify{ratio $\rho_{\mathrm{ave}}$} in the window $25u^{-1} < t < 40u^{-1}$ for several values of $J_{\mathrm{exc}}$,
    with $T_p = 20u^{-1}$ and $L_x = 16a$.
    The error bar on each data point represents the maximum deviation of $\rho(t)$ from the averaged value.
    }
    \label{fig:depJ}
\end{figure}

We next investigate how the transmitted spin current is affected by the exchange coupling parameter $J_{\mathrm{exc}}$ at the interfaces of the TDSM and the FMIs.
Figure \ref{fig:depJ}(a) shows the time evolution of \modify{the ratio $\rho(t)$ between $\tau_{\mathrm{DL}}(t)$ and $\bar{\tau}_{\mathrm{stt}}(t)$} for several values of $J_{\mathrm{exc}}$, with $T_p=20u^{-1}$.
While \modify{$\rho(t)$} for $J_{\mathrm{exc}} \gtrsim 0.4u$ shows a plateau close to \modify{unity} at the early stage of the magnetization dynamics,
the plateau for $J_{\mathrm{exc}} = 0.2u$ is lower than \modify{unity}.
The suppression of the plateau for small $J_{\mathrm{exc}}$ can be clearly seen by plotting the time-averaged value of \modify{$\rho(t)$},
\modify{
    \begin{align}
        \rho_{\mathrm{ave}} = \frac{1}{t_{\mathrm{fin}} - t_{\mathrm{ini}}} \int_{t_{\mathrm{ini}}}^{t_{\mathrm{fin}}} dt \ \rho(t)
    \end{align}    
}
for $t_{\mathrm{ini}} = 25u^{-1}$ and $t_{\mathrm{fin}} = 40u^{-1}$,
which is shown in Fig.~\ref{fig:depJ}(b).
The dependence on $J_{\mathrm{exc}}$ can again be understood from the surface transport picture:
the semi-quantized spin current is achieved if the magnetization dynamics is adiabatic,
which requires the exchange gap $2J_{\mathrm{exc}}$ on the surface spectrum to be much larger than the precession frequency $\Omega = 2\pi/T_p \approx 0.3u^{-1}$ of the magnetization $\boldsymbol{n}_1(t)$ (for $T_p=20u^{-1}$).

\begin{figure}[tbp]
    \begin{center}
        \includegraphics[width=9cm,clip]{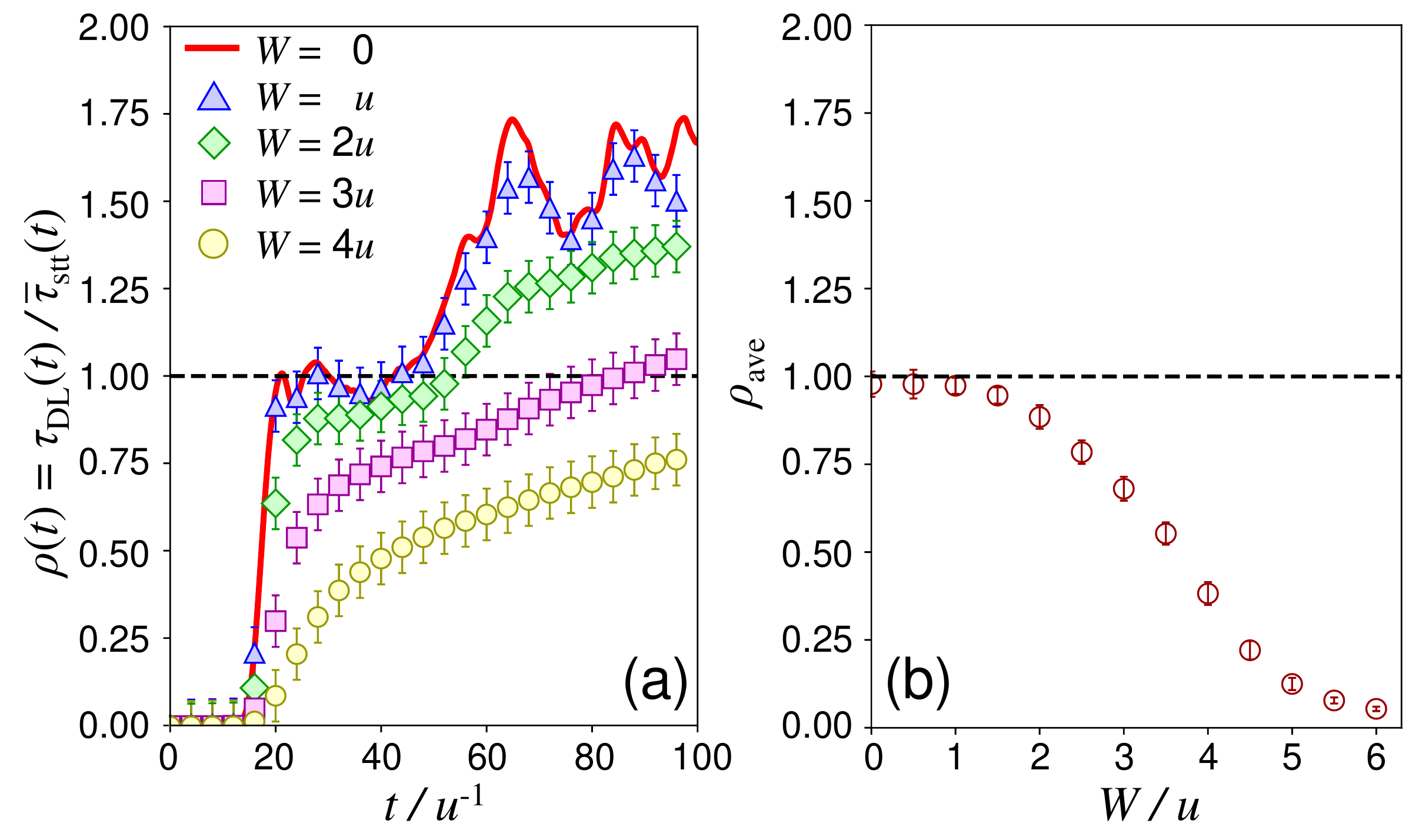}
    \end{center}
    \caption{
    (a) Time evolution of the  \modify{ratio $\rho(t) = \tau_{\mathrm{DL}}(t) / \bar{\tau}_{\mathrm{stt}}(t)$} under the random disorder potential
    for several values of the disorder strength $W$,
    with $J_{\mathrm{exc}} = 0.5u$, $T_p = 20u^{-1}$, and $L_x = 16a$.
    The error bar on each data point represents the standard error of \modify{$\rho(t)$} with respect 
    to the \modify{disorder}.
    (b) The time-averaged \modify{value $\rho_{\mathrm{ave}}$ of the ratio $\rho(t)$} in the window $25u^{-1} < t < 40u^{-1}$ for several values of $W$,
    with $J_{\mathrm{exc}} = 0.5u$, $T_p = 20u^{-1}$, and $L_x = 16a$.
    }
    \label{fig:depW}
\end{figure}
  
In order to check the robustness of spin transmission against \modify{disorder},
we introduce the local random potential
\begin{align}
    \mathcal{H}_{\mathrm{dis}} &= \sum_{\br} V_{\br} c_{\br}^\dag c_{\br},
\end{align}
where $V_{\br}$ takes a random value $V_{\br} \in [-W/2, W/2]$ for each lattice site $\br$,
with $W$ characterizing the strength of the disorder.
With 20 profiles of the random disorder potential $V_{\br}$,
we simulate the time evoltion of $\boldsymbol{n}_2(t)$,
and take average of $\boldsymbol{n}_2(t)$ over the 20 profiles to evaluate the disorder-averaged behavior.
The time evolution of the \modify{ratio $\rho(t) = \tau_{\mathrm{DL}}(t) / \bar{\tau}_{\mathrm{stt}}(t)$} 
and its time-averaged value \modify{$\rho_{\mathrm{ave}}$}
in the window $t = 25u^{-1}$ - $40u^{-1}$
are shown in Fig.~\ref{fig:depW}(a) and (b),
with $T_p = 20u^{-1}$, $J = 0.5u$, and $L_x = 16a$.
We calculate both the standard errors of the dampinglike torque with respect 
to the \modify{disorder} and the disorder-averaged oscillations of the dampinglike torque,
and plot the larger one as the error bar for each data point.
We can see that \modify{the plateau $\rho(t) \approx 1$} is achieved for a weak disorder $W \lesssim 2u$,
due to the robustness of the helical surface states under \modify{disorder} with time-reversal symmetry.
The plateau value is gradually suppressed under a strong disorder,
once its magnitude exceeds the bandwidth $\sim 2u$ of the Dirac bands in the bulk.

\begin{figure}[tbp]
    \begin{center}
        \includegraphics[width=9cm,clip]{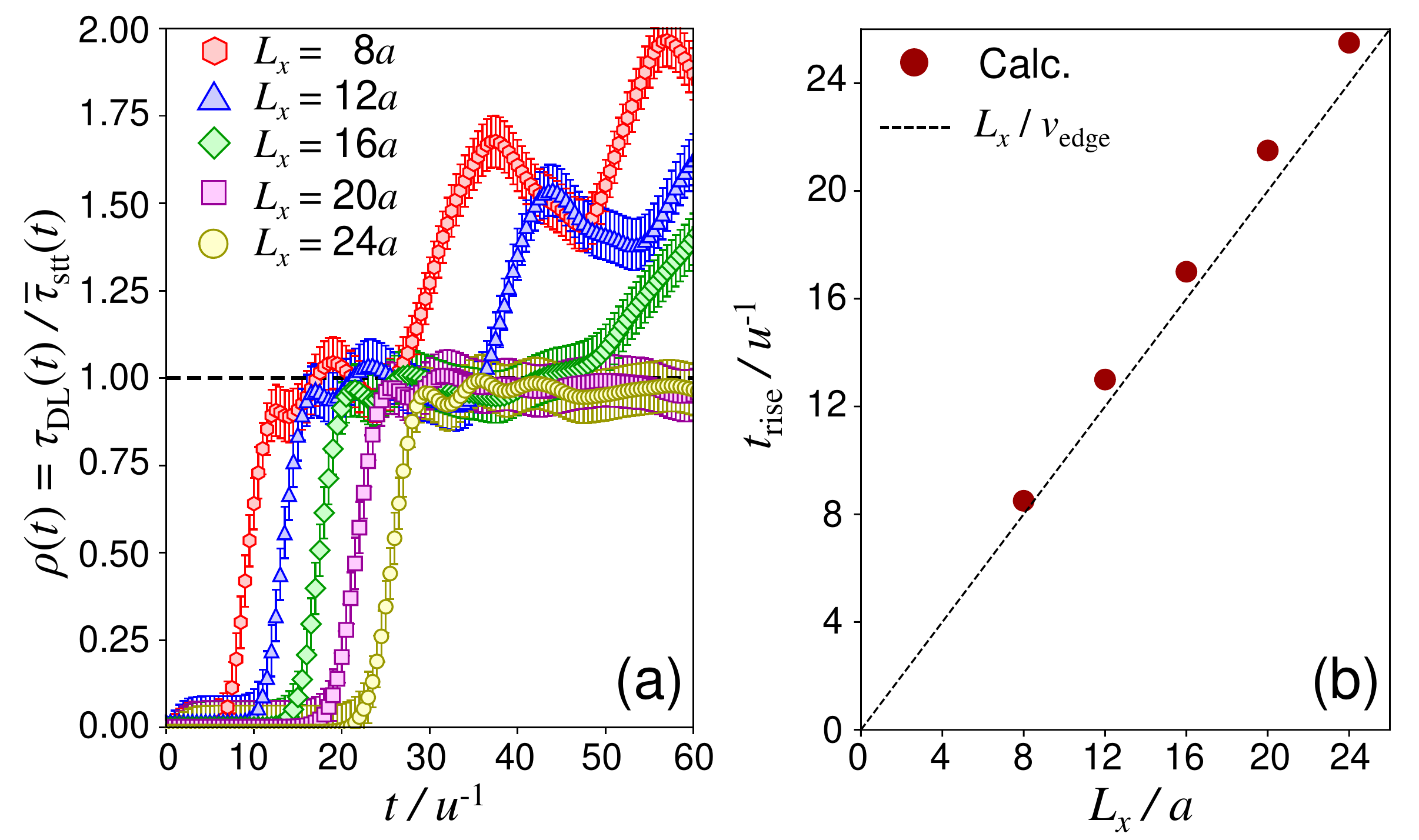}
    \end{center}
    \caption{
    (a) Time evolution of the \modify{ratio $\rho(t) = \tau_{\mathrm{DL}}(t) / \bar{\tau}_{\mathrm{stt}}(t)$} under the random disorder potential
    of $W=u$ for several values of the system size $L_x$,
    with $J_{\mathrm{exc}} = 0.5u$ and $T_p = 20u^{-1}$.
    The error bar on each data point represents the standard error of \modify{$\rho(t)$} with respect 
    to the \modify{disorder}.
    (b) The relation between the system size $L_x$ and the time $t_{\mathrm{rise}}$ when the calculated \modify{ratio $\rho(t) = \tau_{\mathrm{DL}}(t) / \bar{\tau}_{\mathrm{stt}}(t)$} rises to the plateau \modify{$\rho(t) \approx 1$}.
    The calculated $t_{\mathrm{rise}}$ well agrees with $t_{\mathrm{rise}} = L_x/v_{\mathrm{edge}}$,
    which comes from the surface transport picture with the velocity $v_{\mathrm{edge}} = ua$.
    }
    \label{fig:depLx}
\end{figure}

Finally, in order to check the robustness of the surface spin transport over a long range,
we vary the system size $L_x$ and observe its effect on the torques.
In Fig.~\ref{fig:depLx}(a),
we show the time-evolution \modify{of the ratio $\rho(t) = \tau_{\mathrm{DL}}(t) / \bar{\tau}_{\mathrm{stt}}(t)$}
for several values of $L_x$,
with the disorder strength $W=u$.
We can see that \modify{$\rho(t)$ rises to the plateau $\approx 1$}
at $t_{\mathrm{rise}} \approx L_x / v_{\mathrm{edge}}$ (here $v_{\mathrm{edge}} \approx ua$),
as shown in Fig.~\ref{fig:depLx}(b),
namely the time when an electron propagating from FM1 arrives at FM2.
Moreover, the semi-quantized plateau is not significantly violated by the disorder,
even for a long distance $L_x = 24a$.
From these results, we can conclude that the helical surface states of the TDSM
can realize a long-range spin transport that is robust against a moderate disorder below the bulk bandwidth.

\section{Conclusion} \label{sec:conclusion}
In this article, we have theoretically demonstrated a long-range spin transport
realized by the surface states of a TDSM.
TDSMs, such as $\mathrm{Cd_3 As_2}$ and $\mathrm{Na_3 Bi}$,
have quasi-1D gapless states on the surface in the form of Fermi arcs,
which are spin-helical and robust against \modify{disorder} keeping time-reversal symmetry.
By taking a junctions of two FMIs and a TDSM as a model setup, as shown in Fig.~\ref{fig:schem},
we have investigated the spin transfer between the two FMIs driven by the magnetization dynamics in one FMI (FM1).
We have evaluated the spin transfer
both analytically by evaluating the electron numbers in the helical surface channels based on the 1D scattering theory,
and numerically by simulating the real-time evolution of all the electrons in the TDSM on a lattice model.
As a result,
we have found that the spin transfer between the two FMIs at charge neutrality is dominated by the helical surface states,
and that such a surface spin transport is almost insensitive to the \modify{disorder} keeping time-reversal symmetry.

In particular, at the early stage of the spin transmission after turning on the magnetization dynamics,
we have found that the transmitted spin current reaches the semi-quantized value $\bar{j}^s_2$,
which is a universal value determined
by the precession frequency $\Omega$ of the magnetization dynamics in FM1
and the number of helical channels $\nu_z$ on the surface,
corresponding to the distance of the Dirac points $2k_D$ in momentum space.
This semi-quantized spin current is achieved
if the magnetization opens a large exchange gap on the helical surface states in comparison with the frequency $\Omega$.
This condition is in common with the quantized charge pumping
driven by magnetization dynamics on the helical edge states of QSHI \cite{Qi_2008,Chen_2010,Mahfouzi_2010,Hattori_2013,Meng_2014,Deng_2015,Wang_2019,Araki_2020,Misawa_2019}.
Indeed, as discussed in Section \ref{sec:edge-picture},
the (semi-)quantized charge current and spin current are described in the unified framework based on the edge (surface) transport picture,
and they are related independently of the coupling to the FMIs, as in Eq.~(\ref{eq:spin-current}).
Since our analysis and simulation show that the trasmitted spin current will deviate from the semi-quantized value
after a long time of magnetization dynamics in FM1,
we expect that the semi-quantized spin current can be measured if the magnetization dynamics is in a short time,
e.g. driven by a microwave pulse.

From our findings in this article,
we expect that the helical surface states of the TDSM are advantageous for a long-range spin transport,
in comparison with conduction electrons in normal metals or magnons (spin waves) in magnetic insulators.
Plus, in comparison with 1D edge states of 2D topological insulators (QSHIs) and Chern insulators,
the helical surface states of 3D TDSMs are advantageous
in that they consist of many 1D channels and are capable of transferring a large spin current.
The recent transport measurement of a heterostructure of a TDSM $\mathrm{Cd_3 As_2}$ and a FMI
indicates the effect of exchange splitting on the surface states \cite{Uchida_2019},
and hence we may expect that the long-range spin transport can be possibly measured with such heterostructures of $\mathrm{Cd_3 As_2}$.

\acknowledgments{
    This work is supported by JSPS KAKENHI Grant Number 20H01830.
    Y.~A. is supported by the Leading Initiative for Excellent Young Researchers (LEADER).
    T.~M. is supported by JSPS KAKENHI Grants Numbers JP16H06345 and JP19K03739, 
    and by Building of Consortia for the Development of Human Resources in 
    Science and Technology from the MEXT of Japan.
    K.~N. is supported by JST CREST Grant No.~JPMJCR18T2.
}

\appendix

\section{$S$-matrix} \label{sec:S-matrix}

We here evaluate the scattering rates $R(\epsilon)$ and $T(\epsilon)$ introduced in Section \ref{sec:pumping},
which are defined in the rotating frame of spin.
Assuming that the FMI is of length $L$ and located at $x_\parallel=0$,
the Hamiltonian for the edge electrons in the rest frame is given as
\begin{align}
    H(t) &= v_{\mathrm{edge}} p_\parallel \sigma_z + J \boldsymbol{n}(t) \cdot \boldsymbol{\sigma} \pi_L(x_\parallel).
\end{align}
where $\pi_L(x_\parallel)$ is the rectangular function taking a value $1$ for $-L/2<x_\parallel<L/2$ and $0$ otherwise,
and $p_\parallel = -i \partial/\partial x_{\parallel}$ is the momentum operator along $x_\parallel$.
We take the precession of magnetization around $z$-axis,
where the magnetization direction $\boldsymbol{n}(t)$ is given as
\begin{align}
    \boldsymbol{n}(t) =(\sin\theta \cos(\Omega t +\phi) , \sin\theta \sin(\Omega t +\phi) , \cos\theta).
\end{align}
We introduce $J_z = J \cos\theta$ and $J_\perp = J\sin\theta$ for later discussions.
$J_\perp$ gives the exchange gap, if the magnetization is stationary.

\begin{figure}[tbp]
    \includegraphics[width=8.4cm]{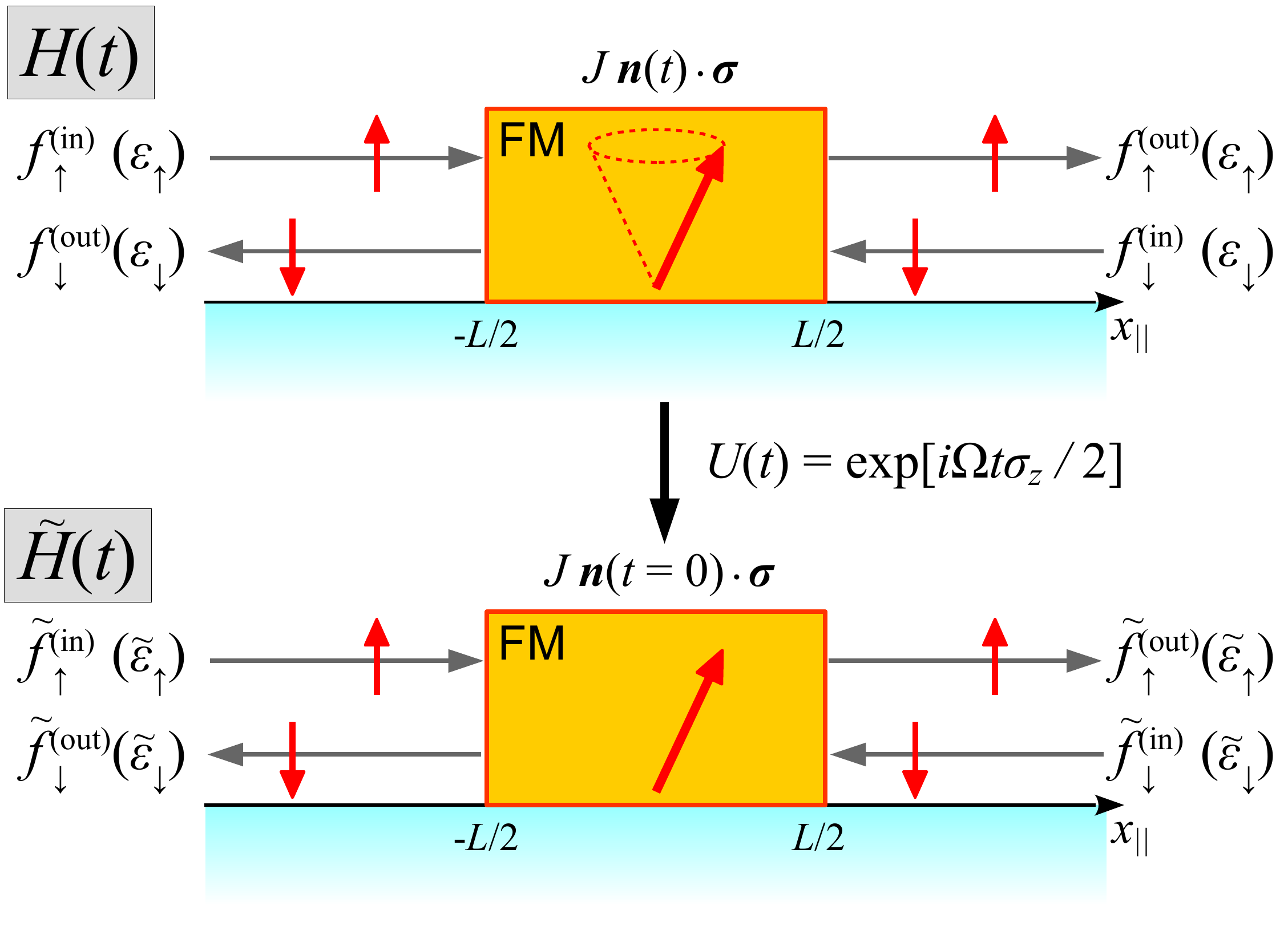}
    \caption{Schematic pictures for the scattering problem in the rest frame (top panel) and the rotating frame (bottom).
    By the unitary transformation $U(t)$ to the rotating frame of spin,
    the direction of the magnetization is fixed and the energies of the incoming and outgoing electrons are shifted.}
    \label{fig:edge-state}
\end{figure}     

The magnetization $\boldsymbol{n}(t)$ is fixed in the rotating frame of spin.
By the time-dependent unitary transformation
\begin{align}
    U(t) = e^{i \Omega t \sigma_z/2},
\end{align}
which rotates spin by the angular velocity $\Omega$ around $z$-axis,
the Hamiltonian becomes time-independent,
\begin{align}
    \tilde{H} &= U(t) \left[H(t) - i\partial_t \right] U^\dag(t) \label{eq:H-rotated} \\
    &= (v_{\mathrm{edge}} p_\parallel - \tfrac{\Omega}{2})\sigma_z + J \boldsymbol{n}(t=0) \cdot \boldsymbol{\sigma} \pi_L(x_\parallel), \nonumber
\end{align}
as schematically shown in Fig.~\ref{fig:edge-state}.
Therefore, we can apply the conventional scattering theory with a time-independent scatterer
in this rotating frame,
to treat the charge and spin pumping by the precessing magnetization.
We here fix the energy $\tilde{\epsilon}$ in this frame,
evaluate the eigenstate in each region, and connect the obtained eigenstates to derive the scattering solution.

(i) In the nonmagnetic region, the solutions are simply the eigenstates of spin.
The plane-wave solutions for right-moving spin-$\uparrow$ electrons and  left-moving spin-$\downarrow$ electrons are given as
\begin{align}
    e^{i k_\uparrow x_\parallel} \begin{pmatrix} 1 \\ 0 \end{pmatrix} & \qquad \left(v_{\mathrm{F}} k_\uparrow = \tilde{\epsilon}+\tfrac{\Omega}{2} \right) \\
    e^{i k_\downarrow x_\parallel} \begin{pmatrix} 0 \\ 1 \end{pmatrix}, & \qquad \left(v_{\mathrm{F}} k_\downarrow = -\tilde{\epsilon}+\tfrac{\Omega}{2} \right)
\end{align}
respectively.

(ii) The solutions in the magnetic region are given as
\begin{align}
    e^{i k_\pm x_\parallel} \phi^\pm_{\tilde{\epsilon}},
\end{align}
where the momentum $k_\pm$ is defined by
\begin{align}
    v_{\mathrm{edge}} k_\pm = \pm v_{\mathrm{edge}} K -J_z +\tfrac{\Omega}{2} \quad
    \left( v_{\mathrm{edge}} K \equiv \sqrt{\tilde{\epsilon}^2-J_\perp^2} \right)
\end{align}
and $\phi^\pm_{\tilde{\epsilon}}$ is the eigenvector of the matrix
\begin{align}
    \tilde{H}|_{p_\parallel = k_\pm, \pi_L(x_\parallel) =1} &=
    \begin{pmatrix}
        v_{\mathrm{edge}} k_\pm -\tfrac{\Omega}{2} + J_z & J_\perp e^{-i\phi} \\
        J_\perp e^{i\phi} & -v_{\mathrm{edge}} k_\pm +\tfrac{\Omega}{2} - J_z
    \end{pmatrix} \nonumber\\
    &=
    \begin{pmatrix}
        \pm v_{\mathrm{edge}} K & J_\perp e^{-i\phi} \\
        J_\perp e^{i\phi} & \mp v_{\mathrm{edge}} K
    \end{pmatrix}
\end{align}
for the eigenvalue $\tilde{\epsilon}$.
In particular, one can write $\phi^\pm_{\tilde{\epsilon}}$ as
\begin{align}
    \phi^+_{\tilde{\epsilon}} = \begin{pmatrix} u_{\tilde{\epsilon}} \\ v_{\tilde{\epsilon}}e^{i\phi} \end{pmatrix}, \quad
    \phi^-_{\tilde{\epsilon}} = \begin{pmatrix} v_{\tilde{\epsilon}}e^{-i\phi} \\ u_{\tilde{\epsilon}} \end{pmatrix},
\end{align}
with
\begin{align}
    \begin{pmatrix} u_{\tilde{\epsilon}} \\ v_{\tilde{\epsilon}} \end{pmatrix}
    =
    \begin{pmatrix} \tilde{\epsilon} + v_{\mathrm{edge}} K \\ J_\perp  \end{pmatrix},
\end{align}
without normalization.
If $\tilde{\epsilon}$ is inside the exchange gap $(|\tilde{\epsilon}|<J)$,
$k_\pm$ becomes imaginary and the solutions become exponentially-growing and decaying functions.

With the above solutions, we can construct the overall wave function as
\begin{align}
    \psi(x_\parallel) = 
    \begin{cases}
        A_\uparrow e^{i k_\uparrow x_\parallel} \left(\begin{smallmatrix} 1 \\ 0 \end{smallmatrix}\right)
        + A_\downarrow e^{i k_\downarrow x_\parallel} \left(\begin{smallmatrix} 0 \\ 1 \end{smallmatrix}\right)
        & (x_\parallel < -\tfrac{L}{2}) \\
        B_+ e^{i k_+ x_\parallel} \phi^+_{\tilde{\epsilon}}
        + B_- e^{i k_- x_\parallel} \phi^-_{\tilde{\epsilon}}
        & (-\tfrac{L}{2} < x_\parallel < \tfrac{L}{2}) \\
        C_\uparrow e^{i k_\uparrow x_\parallel} \left(\begin{smallmatrix} 1 \\ 0 \end{smallmatrix}\right)
        + C_\downarrow e^{i k_\downarrow x_\parallel} \left(\begin{smallmatrix} 0 \\ 1 \end{smallmatrix}\right).
        & (\tfrac{L}{2} < x_\parallel)
    \end{cases}
\end{align}
\begin{widetext}
    The boundary conditions at $x_\parallel = \pm \tfrac{L}{2}$ lead to the relations
    \begin{align}
        A_\uparrow e^{-i k_\uparrow L/2} \begin{pmatrix} 1 \\ 0 \end{pmatrix}
        + A_\downarrow e^{-i k_\downarrow L/2} \begin{pmatrix} 0 \\ 1 \end{pmatrix}
        &= B_+ e^{-i k_+ L/2} \begin{pmatrix} u_{\tilde{\epsilon}} \\ v_{\tilde{\epsilon}} e^{i\phi} \end{pmatrix}
        + B_- e^{-i k_- L/2} \begin{pmatrix} v_{\tilde{\epsilon}} e^{-i\phi} \\ u_{\tilde{\epsilon}} \end{pmatrix} \\
        C_\uparrow e^{i k_\uparrow L/2} \begin{pmatrix} 1 \\ 0 \end{pmatrix}
        + C_\downarrow e^{i k_\downarrow L/2} \begin{pmatrix} 0 \\ 1 \end{pmatrix}
        &= B_+ e^{i k_+ L/2} \begin{pmatrix} u_{\tilde{\epsilon}} \\ v_{\tilde{\epsilon}} e^{i\phi} \end{pmatrix}
        + B_- e^{i k_- L/2} \begin{pmatrix} v_{\tilde{\epsilon}} e^{-i\phi} \\ u_{\tilde{\epsilon}} \end{pmatrix}
    \end{align}
    which can be assembled into the matrix forms as
    \begin{align}
        \begin{pmatrix}
            A_\uparrow e^{-i k_\uparrow L/2}  \\
            A_\downarrow e^{-i k_\downarrow L/2} 
        \end{pmatrix}
        &=
        \begin{pmatrix}
            u_{\tilde{\epsilon}} e^{-i k_+ L/2} & v_{\tilde{\epsilon}} e^{-i\phi-i k_- L/2} \\
            v_{\tilde{\epsilon}} e^{i\phi -i k_+ L/2} & u_{\tilde{\epsilon}} e^{-i k_- L/2}
        \end{pmatrix}
        \begin{pmatrix} B_+ \\ B_- \end{pmatrix} \\
        \begin{pmatrix}
            C_\uparrow e^{i k_\uparrow L/2}  \\
            C_\downarrow e^{i k_\downarrow L/2} 
        \end{pmatrix}
        &=
        \begin{pmatrix}
            u_{\tilde{\epsilon}} e^{i k_+ L/2} & v_{\tilde{\epsilon}} e^{-i\phi+i k_- L/2} \\
            v_{\tilde{\epsilon}} e^{i\phi +i k_+ L/2} & u_{\tilde{\epsilon}} e^{i k_- L/2}
        \end{pmatrix}
        \begin{pmatrix} B_+ \\ B_- \end{pmatrix}.
    \end{align}
    Therefore, the relation between $A_{\uparrow/\downarrow}$ and $C_{\uparrow/\downarrow}$ is given as
    \begin{align}
        \begin{pmatrix}
            C_\uparrow e^{i k_\uparrow L/2}  \\
            C_\downarrow e^{i k_\downarrow L/2} 
        \end{pmatrix}
        &= 
        \begin{pmatrix}
            u_{\tilde{\epsilon}} e^{i k_+ L/2} & v_{\tilde{\epsilon}} e^{-i\phi+i k_- L/2} \\
            v_{\tilde{\epsilon}} e^{i\phi +i k_+ L/2} & u_{\tilde{\epsilon}} e^{i k_- L/2}
        \end{pmatrix}
        \begin{pmatrix}
            u_{\tilde{\epsilon}} e^{-i k_+ L/2} & v_{\tilde{\epsilon}} e^{-i\phi-i k_- L/2} \\
            v_{\tilde{\epsilon}} e^{i\phi -i k_+ L/2} & u_{\tilde{\epsilon}} e^{-i k_- L/2}
        \end{pmatrix}^{-1}
        \begin{pmatrix}
            A_\uparrow e^{-i k_\uparrow L/2}  \\
            A_\downarrow e^{-i k_\downarrow L/2} 
        \end{pmatrix}
        \\
        &=
        \frac{1}{u_{\tilde{\epsilon}}^2 - v_{\tilde{\epsilon}}^2}
        \begin{pmatrix}
            u_{\tilde{\epsilon}}^2 e^{i k_+ L} - v_{\tilde{\epsilon}}^2 e^{i k_- L} & u_{\tilde{\epsilon}}v_{\tilde{\epsilon}} e^{-i\phi} (e^{i k_- L} - e^{i k_+ L}) \\
            u_{\tilde{\epsilon}}v_{\tilde{\epsilon}} e^{i\phi} (e^{i k_+ L} - e^{i k_- L}) & u_{\tilde{\epsilon}}^2 e^{i k_- L} - v_{\tilde{\epsilon}}^2 e^{i k_+ L}
        \end{pmatrix}
        \begin{pmatrix}
            A_\uparrow e^{-i k_\uparrow L/2}  \\
            A_\downarrow e^{-i k_\downarrow L/2} 
        \end{pmatrix}.
    \end{align}
    From the definitions $v_{\mathrm{edge}} k_{\uparrow/\downarrow} = \pm \tilde{\epsilon} +\tfrac{\Omega}{2}$ and $v_{\mathrm{edge}} k_\pm = \pm v_{\mathrm{edge}} K -J_z + \tfrac{\Omega}{2}$,
    this relation further reduces as
    \begin{align}
        \begin{pmatrix}
            C_\uparrow e^{i \tilde{\epsilon} L/2v_{\mathrm{edge}}}  \\
            C_\downarrow e^{-i \tilde{\epsilon} L/2v_{\mathrm{edge}}} 
        \end{pmatrix}
        &=
        \frac{e^{-i J_z L/v_{\mathrm{edge}}}}{u_{\tilde{\epsilon}}^2 - v_{\tilde{\epsilon}}^2}
        \begin{pmatrix}
            u_{\tilde{\epsilon}}^2 e^{i K L} - v_{\tilde{\epsilon}}^2 e^{-i K L} & u_{\tilde{\epsilon}}v_{\tilde{\epsilon}} e^{-i\phi} (e^{-i K L} - e^{i K L}) \\
            u_{\tilde{\epsilon}}v_{\tilde{\epsilon}} e^{i\phi} (e^{i K L} - e^{-i K L}) & u_{\tilde{\epsilon}}^2 e^{-i K L} - v_{\tilde{\epsilon}}^2 e^{i K L}
        \end{pmatrix}
        \begin{pmatrix}
            A_\uparrow e^{-i\tilde{\epsilon} L/2v_{\mathrm{edge}}}  \\
            A_\downarrow e^{i \tilde{\epsilon} L/2v_{\mathrm{edge}}} 
        \end{pmatrix} \\
        &\equiv e^{-i J_z L/v_{\mathrm{edge}}} 
        \begin{pmatrix}
            \Lambda_{\uparrow\uparrow} & \Lambda_{\uparrow\downarrow} \\
            \Lambda_{\downarrow\uparrow} & \Lambda_{\downarrow\downarrow}
        \end{pmatrix}
        \begin{pmatrix}
            A_\uparrow e^{-i\tilde{\epsilon} L/2v_{\mathrm{edge}}}  \\
            A_\downarrow e^{i \tilde{\epsilon} L/2v_{\mathrm{edge}}} 
        \end{pmatrix}
        .
    \end{align}
\end{widetext}

Here we need to recast the above relation into the form of $S$-matrix,
\begin{align}
    \begin{pmatrix}
        C_\uparrow \\ A_\downarrow 
    \end{pmatrix}
    =
    \begin{pmatrix}
        \tilde{t}_{\uparrow\uparrow}(\tilde{\epsilon}) & \tilde{r}_{\uparrow\downarrow}(\tilde{\epsilon}) \\
        \tilde{r}_{\downarrow\uparrow}(\tilde{\epsilon}) & \tilde{t}_{\downarrow\downarrow}(\tilde{\epsilon})
    \end{pmatrix}
    \begin{pmatrix}
        A_\uparrow \\ C_\downarrow
    \end{pmatrix}.
\end{align}
The relation
\begin{align}
    C_\downarrow e^{-i \frac{\tilde{\epsilon} L}{2v_{\mathrm{edge}}}} = e^{-i \frac{J_z L}{v_{\mathrm{edge}}}} \left[ \Lambda_{\downarrow\uparrow} A_\uparrow e^{-i\frac{\tilde{\epsilon} L}{2v_{\mathrm{edge}}}} + \Lambda_{\downarrow\downarrow}A_\downarrow e^{i \frac{\tilde{\epsilon} L}{2v_{\mathrm{edge}}}} \right]
\end{align}
can be rewritten as
\begin{align}
    A_\downarrow &= e^{-i\frac{\tilde{\epsilon}}{v_{\mathrm{edge}}}L} \left[ -\frac{\Lambda_{\downarrow\uparrow}}{\Lambda_{\downarrow\downarrow}}  A_\uparrow + \frac{e^{i \frac{J_z}{v_{\mathrm{edge}}}L} }{\Lambda_{\downarrow\downarrow}} C_\downarrow \right],
\end{align}
which yields 
\begin{align}
    \tilde{r}_{\downarrow\uparrow}(\tilde{\epsilon}) &= - e^{-i\frac{\tilde{\epsilon}}{v_{\mathrm{edge}}}L}  \frac{\Lambda_{\downarrow\uparrow}}{\Lambda_{\downarrow\downarrow}}\\
    &= - e^{-i(\frac{\tilde{\epsilon}}{v_{\mathrm{edge}}}L -\phi)} \frac{u_{\tilde{\epsilon}}v_{\tilde{\epsilon}} (e^{i K L} - e^{-i K L})}{u_{\tilde{\epsilon}}^2 e^{-i K L} - v_{\tilde{\epsilon}}^2 e^{i K L}} \\
    \tilde{t}_{\downarrow\downarrow}(\tilde{\epsilon}) &= e^{-i\frac{\tilde{\epsilon}}{v_{\mathrm{edge}}}L} \frac{e^{i \frac{J_z}{v_{\mathrm{edge}}}L} }{\Lambda_{\downarrow\downarrow}} \\
    &= e^{-i\frac{\tilde{\epsilon}-J_z}{v_{\mathrm{edge}}} L}  \frac{u_{\tilde{\epsilon}}^2 - v_{\tilde{\epsilon}}^2 }{u_{\tilde{\epsilon}}^2 e^{-i K L} - v_{\tilde{\epsilon}}^2 e^{i K L}}.
\end{align}
The relation
\begin{align}
    C_\uparrow e^{i \frac{\tilde{\epsilon} L}{2v_{\mathrm{edge}}}} = e^{-i \frac{J_z L}{v_{\mathrm{edge}}}} \left[ \Lambda_{\uparrow\uparrow} A_\uparrow e^{-i\frac{\tilde{\epsilon} L}{2v_{\mathrm{edge}}}} + \Lambda_{\uparrow\downarrow}A_\downarrow e^{i \frac{\tilde{\epsilon} L}{2v_{\mathrm{edge}}}} \right]
\end{align}
can be rewritten as
\begin{align}
    C_\uparrow &= e^{-i \frac{J_z}{v_{\mathrm{edge}}}L} \left[ \Lambda_{\uparrow\uparrow} A_\uparrow e^{-i\frac{\tilde{\epsilon}}{v_{\mathrm{edge}}} L} + \Lambda_{\uparrow\downarrow}A_\downarrow \right] \\
    &= e^{-i \frac{J_z}{v_{\mathrm{edge}}}L} \left[ \Lambda_{\uparrow\uparrow} A_\uparrow e^{-i\frac{\tilde{\epsilon}}{v_{\mathrm{edge}}} L} + \Lambda_{\uparrow\downarrow}(\tilde{r}_{\downarrow\uparrow}A_\uparrow + \tilde{t}_{\downarrow\downarrow}C_\downarrow ) \right],
\end{align}
which yields
\begin{align}
    \tilde{t}_{\uparrow\uparrow}(\tilde{\epsilon}) &= e^{-i \frac{J_z}{v_{\mathrm{edge}}}L} \left[ \Lambda_{\uparrow\uparrow}e^{-i\frac{\tilde{\epsilon}}{v_{\mathrm{edge}}} L} + \Lambda_{\uparrow\downarrow}\tilde{r}_{\downarrow\uparrow}(\tilde{\epsilon}) \right] \\
    &= e^{-i\frac{\tilde{\epsilon}+J_z}{v_{\mathrm{edge}}} L}  \frac{u_{\tilde{\epsilon}}^2 - v_{\tilde{\epsilon}}^2 }{u_{\tilde{\epsilon}}^2 e^{-i K L} - v_{\tilde{\epsilon}}^2 e^{i K L}} \\
    \tilde{r}_{\uparrow\downarrow}(\tilde{\epsilon}) &= e^{-i \frac{J_z}{v_{\mathrm{edge}}}L} \Lambda_{\uparrow\downarrow} \tilde{t}_{\downarrow\downarrow}(\tilde{\epsilon}) \\
    &= - e^{-i(\frac{\tilde{\epsilon}}{v_{\mathrm{edge}}}L +\phi)} \frac{u_{\tilde{\epsilon}}v_{\tilde{\epsilon}} (e^{i K L} - e^{-i K L})}{u_{\tilde{\epsilon}}^2 e^{-i K L} - v_{\tilde{\epsilon}}^2 e^{i K L}}.
\end{align}

The obtained reflection and transmission rates in the rotated frame satisfy the detailed balance relations
\begin{align}
    |\tilde{r}_{\uparrow\downarrow}(\tilde{\epsilon})|^2 &= |\tilde{r}_{\downarrow\uparrow}(\tilde{\epsilon})|^2 \\
    |\tilde{t}_{\uparrow\uparrow}(\tilde{\epsilon})|^2 &= |\tilde{t}_{\downarrow\downarrow}(\tilde{\epsilon})|^2 ,
\end{align}
which are $R(\tilde{\epsilon})$ and $T(\tilde{\epsilon})$ defined in the main text.
From the above calculations, these rates are explicitly obtained as
\begin{align}
    R(\tilde{\epsilon}) &= \frac{1 - \cos (2KL)}{(2\tilde{\epsilon}^2/J_\perp^2) -1 -\cos(2KL)} \\
    T(\tilde{\epsilon}) &= \frac{(2\tilde{\epsilon}^2/J_\perp^2) -2}{(2\tilde{\epsilon}^2/J_\perp^2) -1 -\cos(2KL)}, 
\end{align}
where
\begin{align}
    K = \frac{1}{v_{\mathrm{edge}}} \sqrt{\tilde{\epsilon}^2 -J_\perp^2}.
\end{align}
We can check that they satisfy the unitarity condition
\begin{align}
    R(\tilde{\epsilon}) + T(\tilde{\epsilon}) =1.
\end{align}

\end{document}